\documentclass[usenatbib]{mn2e}
\usepackage{amssymb, latexsym, amsmath, graphicx, url}

\title{The duty cycle of radio-mode feedback in complete samples of clusters}
\author[L.~B\^{\i}rzan et al.]{L.~B\^{\i}rzan,$^{1}$ D.~A.~Rafferty,$^{1}$ P.~E.~J.~Nulsen,$^{2}$ B.~R.~McNamara,$^{2,3,4}$
\newauthor
H.~J.~A.~R\"{o}ttgering$^{1}$ M.~W.~Wise,$^{5}$ and  R.~Mittal$^{6}$\\
$^{1}$ Leiden Observatory, Leiden University, Oort Gebouw, P.O. Box 9513, 2300 RA Leiden, The Netherlands\\
$^{2}$ Harvard-Smithsonian Centre for Astrophysics, 60 Garden St., Cambridge, MA 02138, USA\\
$^{3}$ Department of Physics and Astronomy, University of Waterloo, Waterloo, ON N2L 2G1, Canada\\
$^{4}$ Perimeter Institute for Theoretical Physics, Waterloo, ON N2L 2Y5, Canada\\
$^{5}$ Netherlands Institute for Radio Astronomy (ASTRON), P.O. Box 2, 7990 AA Dwingeloo, The Netherlands\\
$^{6}$ Chester F. Carlson centre for Imaging Science, Rochester Institute of Technology, Rochester, NY 14623, USA}
\begin{document}

\maketitle

\begin{abstract}
The \emph{Chandra} X-ray Observatory has revealed X-ray bubbles in the intracluster medium (ICM) of many nearby cooling flow clusters. The bubbles trace feedback that is thought to couple the central active galactic nucleus (AGN) to the ICM, helping to stabilize cooling flows and govern the evolution of massive galaxies. However, the prevalence and duty cycle of such AGN outbursts is not well understood. To this end, we  study how cooling is balanced by bubble heating for complete samples of clusters (the Brightest 55 clusters of galaxies, hereafter B55, and the HIghest X-ray FLUx Galaxy Cluster Sample, HIFLUGCS). We find that the radio luminosity of the central galaxy only exceeds 2.5 $\times$ 10$^{30}$ erg s$^{-1}$ Hz$^{-1}$ in cooling flow clusters. This result implies a connection between the central radio source and the ICM, as expected if AGN feedback is operating. Additionally, we find a duty cycle for radio mode feedback, the fraction of time that a system possesses bubbles inflated by its central radio source, of  $\gtrsim 69 \%$ for B55 and $\gtrsim 63 \%$ for HIFLUGCS. These duty cycles are lower limits since some bubbles are likely missed in existing images. We used simulations to constrain the bubble power that might be present and remain undetected in the cooling flow systems without detected bubbles. Among theses systems, almost all could have significant bubble power. Therefore, our results imply that the duty cycle of AGN outbursts with the potential to heat the gas significantly in cooling flow clusters is at least 60 per cent and could approach 100 per cent. 
\end{abstract}


\section{Introduction}\label{intro}

A much debated topic in extragalactic astronomy in the last 20 years is the ``cooling flow'' problem \citep{fab94}. Since the cooling time of the ICM in the cores of many galaxy clusters is shorter than a few Gyrs \citep{edge92}, in the absence of heating the gas is expected to cool down below X-ray temperatures  and condense \citep{fab94}. However, searches for evidence of the high mass depositions rates expected in the standard cooling flow model were unsuccessful \citep[e.g.,][]{mcna89}. Data for such clusters from the \emph{Chandra} and  \emph{XMM-Newton} X-ray observatories do not show the signatures of gas cooling below $\sim2$~keV  \citep[e.g.,][]{pete01}. Therefore, it appears that only a small fraction of the expected cooling is occurring in most cooling flows, suggesting some heating mechanism is operating in these systems.

A possible source of heating has been identified: high resolution images from the \emph{Chandra X-ray Observatory} have revealed X-ray bubbles in the ICM of many nearby clusters, groups and ellipticals  (very few systems  were previously known to host bubbles in their atmospheres, e.g, Perseus, Cygnus A).
Notable examples include Perseus \citep{fabi00}, A2052 \citep{blan01} and MS 0735.6+7421 \citep{mcna05}, among many others that have been discovered. In a study of a large sample of such bubble systems, it was found that a central radio source was present in every system, with the radio plasma often filling the bubbles \citep{birz08}. The X-ray bubbles are therefore interpreted as regions where the X-ray emitting gas has been displaced by radio plasma produced by energetic outflows from the central AGN and show that the AGN is injecting significant power into the ICM. The energy that the bubbles inject into the ICM may be important for balancing cooling \citep{mcna00,fabi00,blan01} and inhibiting star formation in massive galaxies \citep{crot06}. 

A number of authors \citep{birz04,dunn04,dunn05,raff06} have analyzed samples of cooling flow clusters with visible bubbles in their environments. Using a sample of such systems, \citet{raff06} found that at least 50\% of the systems have enough energy in their bubbles to balance cooling, considering only the enthalpy, although the integrated energy is sufficient to offset cooling in all systems \citep{mcna07}. The enthalpy is only a lower limit for the total energy injected, as \emph{Chandra} images reveal evidence for the existence of weak shocks \citep{mcna05,fabi06,form07,wise07} with energies comparable to the bubble enthalpies. Additionally, the energy may still be underestimated due to further adiabatic losses, undetected bubbles \citep{nuss06,binn07} or cosmic ray losses \citep{math08}. 

However, it is not clear whether the bubbles are always present when heating is required. Currently, the bubbles are primarily detected by eye. Some of them have bright rims \citep[e.g., A2052;][]{blan01} which make them "clear" cases. However, the detectability of a bubble depends on its location, orientation \citep{enss02,dieh08,brug09}, its angular size and the depth of the observation. A majority of observed bubbles come from systems where the angle between the jet axis and the line of sight is between 45$^{\circ}$ and 90$^{\circ}$ \citep{enss02,brug09}. As a result, we are likely missing some bubbles in the existing images. 

Indeed, many cooling flow systems, which presumably require heating, do not show bubbles in their existing \emph{Chandra} images \citep[e.g, A1650, A2244;][]{dona05}. It is therefore important to understand the biases and selection effects in the detectability of current X-ray bubble samples. However, images of many of these systems are shallow. In these systems, the question is whether there might exist enough undetected bubble power to balance heating.  Also, there might be alternative scenarios for the lack of bubbles in some cooling flow systems, such as the presence of another heating mechanism  \citep[e.g., ``sloshing'';][]{zuho10}, or perhaps because some systems are in a cooling stage, with high rates of star formation, and might not need heating right now \citep[e.g., A1068;][]{mcna04}. 

These issues are important in understanding the duty cycle of AGN heating, which we define to be the fraction of time that a system possesses bubbles inflated by its central radio source (and hence shows clear evidence of AGN heating). The duty cycle gives insight into the heating process and is useful in constraining (or as an input to) simulations of galaxy and cluster formation and evolution that include AGN heating.

Previous work to understand the duty cycle of AGN heating in clusters has been done by \citet{dunn06}, who found an AGN duty cycle of 70\% for cooling flows in the B55 sample. However, their selection of cooling flow systems is based on ROSAT data, which are less sensitive to the presence of a cooling flow than higher-resolution data. For ellipticals the results are more controversial. \citet{dong10} found a duty cycle for AGN feedback of 50\% for a sample of 51 ellipticals. This duty cycle is similar to the \citet{dunn10} finding of 9 out of 18 ellipticals having bubbles, but much higher than \citet{nuls09} finding by using 104 ellipticals from \emph{Chandra} archive, where only 1/4 have bubbles (note however that Nulsen et al.\ found that the duty cycle varies with the X-ray luminosity, so the above results may be consistent).  In this paper we address the question of duty cycle of AGN feedback in complete subsamples of cooling flow clusters using simulations to place limits on AGN heating for the clusters which do not show bubbles in the existing \emph{Chandra} images. For this purpose we chose the B55 \citep{edge90} and HIFLUGCS \citep{reip02} samples, since both samples have already been observed with \emph{Chandra}.  We assume $H_{0}=70$ km s$^{-1}$Mpc$^{-1}$, $\Omega_{\Lambda}=0.7$, and $\Omega_{\rm{M}}=0.3$ throughout.

\section{Sample Definition and Properties}
\subsection{Complete Samples}\label{comp_samples}

In this work we present an analysis of two complete samples: the B55 sample \citep{edge90} and the HIFLUGCS sample \citep{reip02}. The B55 sample is a 2-10 keV flux limited sample of 55 systems (or 56 since we count both A3391 and A3395s) selected using data from the HEAO-1 and Ariel-V surveys \citep{edge90}. HIFLUGCS is a 0.1-2.4 keV flux limited sample of 64 systems based on ROSAT and ASCA observations at galactic latitude  $b > 20 ^{\circ}$. The two complete samples have 44 systems in common. For the systems that belong only to the B55 sample, the majority of them are low latitude systems ($b < 20 ^{\circ}$). These are: AWM7, Perseus, Ophiuchus, A2319, Triungulum Aus, 3C129.1, PKS 0745-191, Cygnus A and A644. M87 is also present only in the B55 sample, since it was excluded from HIFLUGCS based on the fact that the irregular X-ray background and the extended X-ray emission at low latitudes makes the detection of clusters more difficult. A1689 and A3532 are only in the B55 sample due to differences in the X-ray band used to construct the samples. The systems that belong only to the HIFLUGCS sample are in general smaller, cooler ones (e.g., NGC 507, NGC 1399, NGC 1550, EXO 0423.4-0840, MKW 4 etc), and softer-spectrum systems (e.g., Sersic 159/03, A2589, A2657 etc). 

\citet{sun07} showed that there is a class of systems having truncated atmospheres of low entropy gas, with short cooling time, embedded in much more extensive atmospheres of higher entropy gas. Some of these systems are present in our sample and we refer to them as corona systems (they are marked with an asterisk in Table \ref{bubble_prop}, Table \ref{sim_table} and Table \ref{radio_info_table}).

The B55 sample was studied by \citet{dunn06} in order to determine the fraction of cooling flow clusters with bubbles. In order to separate the cooling flow clusters from the non-cooling flow clusters, \citet{dunn06} used ROSAT data. They defined a cluster to be a cooling flow cluster if $t_{\rm{cool}}<3 \times  10^{9}$ yr in the core and $T_{\rm{outer}}/T_{\rm{center}}>2$. They found that 14 out of 20 cooling flow clusters have clear bubbles. Also, \citet{ross10} studied the B55 sample and separated cooling flow from the non-cooling flow clusters using the pseudo-entropy ratio. The HIFLUGCS sample was studied by \citet{sander06, sander09}, who used a temperature ratio in order to separate the cooling flow clusters; \citet{chen07}, who used a mass deposition rate ratio; and \citet{huds10}, who used the cooling time. However, different  definitions of separation give different  answers  for the cooling flow fraction \citep[for a review see][]{sun12}, which might be also affected by the sample bias \citep{ecke11}.

In Table \ref{info_table}, Table \ref{tcool_table}  and Table \ref{radio_info_table} we list all systems from both complete samples (B55 and HIFLUGCS) in order of right accession. Table \ref{info_table} gives the X-ray peak and optical coordinates for the brightest cluster galaxies (BCGs), together with the BCG names and the separation between the ICM's X-ray peak and the BCG center. The coordinates of the X-ray peak are well determined (within 1--2\arcsec) for systems that have a surface brightness peak; for the fainter and more diffuse systems the determination of the separation becomes harder, and we mark these objects with an asterisk in Table \ref{info_table}. The optical coordinates for the BCGs are robustly determined (typically, within 1--2\arcsec), except in a few cases: A2163, A1689 (the optical image is of poor resolution); A1367 which has 2 BCGs, NGC 3862 and NGC 3842, both of them far from the X-ray center, but neither of them appear to be related to the central X-ray emission; and A2256 which has 2 comparable BCGs, NGC 6331 and UGC 10710, plus a few smaller ones (we choose NGC 6331). However, for some ACIS-I observations, the cluster cores are located in the CCD gaps, which affects the determination of the cluster centre (e.g., A399, A3158, A3562, A2255).

\subsection{Bubble Systems}\label{Cav_sect} 
There are 22 and 26 bubble systems that are detected in the B55 and HIFLUGCS samples, respectively. There are, in total, 31 systems with bubbles. Among the 31 systems, the bubble identifications for 24 systems are published in the literature: A2204 \citep{sand09}, Fornax \citep{shur08}, NGC 4636 \citep{bald09}, NGC 5044 \citep{davi10}, and 20 systems in the \citet{raff06} sample of clusters with bubbles. For NGC 507, III Zw 54, NGC 1550, A496, A3391, A1060 and A3532 we identified new bubbles by inspecting the \emph{Chandra} data. For NGC 507, NGC 1550 and A496 more than one \emph{Chandra} observation is available in the archive. In these cases, we used the \emph{ciao} tool \emph{merge-all} in order to add multiple observation for visual inspection of bubbles (see Table~\ref{bubble_prop}). 

The bubbles were identified and measured using unsharp-mask images made from the exposure-corrected images. For each system, a variety of different smoothing scales was investigated. The scales for which the contrast between the bubbles and the surroundings was the greatest were used for measurement. 
However, in some systems (e.g., III Zw 54, NGC 1550), the unsharp-mask image might reveal structure that is not related to bubbles. Thus, the unsharp-mask tool cannot be blindly used for bubble detection. Rather, deeper X-ray and radio images are needed to confirm the presence of marginally detected bubbles. As a result, some bubbles (e.g., III Zw 54, A1060, A3391, A3532) need to be further confirmed by deeper X-ray data. 

Since the bubbles are presumably created by radio lobes, we investigated the presence of a extended radio emission for these systems. The corona systems A3391 and A3532 have large scale radio lobes \citep{mauc03, vent01} much larger than the cavities we measured from the unsharp mask images. On the other hand, the corona system A1060 has very small radio lobes \citep{lind85} that fit well inside the cavities that we measured, and for the corona system III~Zw~54 there is only a NVSS image available in which any lobe emission is not resolved. For the remaining systems, NGC~507 and NGC~1550 have large-scale radio lobes \citep{murg11, dunn10}, but A496 doesn't have any radio image in the literature besides the NVSS one. As in the case of corona systems A3391 and A3532, for NGC~507 the radio lobes are much larger than the cavity we measure, but for NGC~1550 the radio emission fills very well the cavity that we measure. 

As we underline above, for some systems there seems to be a discrepancy between the cavities sizes derived from unsharp-mask images and the extent of the radio lobe emission. For the corona class systems, A3391 and A3532, it might be that the X-ray surface brightness contrast is too low and we are not able to measure the full extent of the cavities. In this case we might have underestimated the bubble power (enthalpy). For NGC 507, the unsharp mask shows a clear small depression in the west side of the cluster and this is the bubble that we list in Table \ref{bubble_prop}. Additionally, there appear to be shallow deficits at the location of the radio lobes \citep{parm86,deRu86,murg11,giac11b}, such as at the northern part of the western radio lobe, but deeper X-ray data is needed to confirm this. It might be that such cavities have a flattened ``pancake'' shape and are filled with entrained material, rather than the perfect void of material and spherical shape that we assumed when we did the calculation, and as a result the contrast between the cavities and the ICM is weakened. This is in agreement with results from \citet{murg11} where it was found that this system has old lobes, and as a result they might have entrained  material, expanded and leaked. As a result, if for NGC 507 one uses the radio lobes sizes and spherical geometry assumption, the bubble power might be overestimated by a large factor.

Bubble radii and distances from the cluster centers are listed Table \ref{bubble_prop}. Also, Table \ref{bubble_prop} lists the buoyancy time, the time required for the bubble to rise buoyantly from the cluster centre to its present location at its terminal velocity, $v_{\rm{t}}$ \citep[for a description of the analysis see][]{birz04}:
\begin{equation}
t_{\rm{buoy}}=R/v_{\rm{t}}\sim R\sqrt{SC/2gV},
\end{equation}
where $R$ is the projected radial distance from the cluster core to the bubble's center, $S$ is the cross section of the bubble, $V$ is the volume of the bubble and the gravitational acceleration $g \approx 2\sigma^2/R$ \citep{binn87}. For the velocity dispersion ($\sigma$) the HyperLeda database\footnote{http://leda.univ-lyon.fr} was used \citep{patu03}; when no measurement of the velocity dispersion was available, $\sigma=281$ km s$^{-1}$ \citep{birz04} was adopted, which was the average value.

\begin{table*}
\begin{minipage}{126mm}
\caption{Bubble Properties}
\label{bubble_prop}
\begin{tabular}{@{}lcccccccc} 
\hline
 &  $t$$^b$ &  $\sigma$$^c$ &  &  $a$$^d$ &  $b$$^e$ &  $R$$^f$ &  $pV$ &  $t_{\rm{buoy}}$ \\
System$^a$ &  (ks) &  (km s$^{-1}$) &  Bubble &  (kpc) &  (kpc) &  (kpc) &  (10$^{58}$ erg) &  (10$^{7}$ yr) \\
\hline
NGC 507    & 70    &  306     & W & 3.7  & 2.6  & 6.9  & 0.018    & 1.1 \\
III Zw 54*  & 21.2  & \ldots  & W & 3.2  & 3.2  & 7.5  & 0.050    & 1.5  \\  
    &  &   & E & 4.8  & 4.1  & 16.9 & 0.047    & 4.0  \\
NGC 1550   & 90    & 308     & W & 4.4  & 3.9  & 8.6  & 0.17     & 1.4 \\ 
A496       & 57.5  & 273     & S & 8.4  & 3.9  & 8.1  & 0.84     & 1.0  \\
    &  &  & N & 9.4  & 5.5  & 12.0 & 1.50     & 2.9  \\
A3391*     & 16.0  & \ldots  & N & 18.8 & 18.8 & 50.3 & 7.30     & 11.0 \\
A1060*     & 31.9  & 185      & N & 0.4  & 0.4  & 0.4  & 0.000055 & 0.05 \\ 
    & &  & S & 0.4  & 0.4  & 0.4  & 0.000048 & 0.05 \\
A3532*     & 9.4   & \ldots   & N & 7.7  & 7.7  & 17.6 & 0.36     & 3.4    \\
    &  &  & S & 5.9  & 5.9  & 10.0 & 0.16     & 1.6  \\
\hline
\end{tabular}
$^a$The systems for which we identified possible bubbles. 
The asterisk denotes the corona class systems \citep{sun07,sun09}.\\
$^b$For III Zw 54, A496, A3391, A1060, A3532, the time on source after re-processing the data (the same as in the Table \ref{tcool_table}). For NGC 507 and NGC 1550 the total time from multiple observations (Obs ID 5801 for NGC 1550).\\
$^c$For the velocity dispersion, the HyperLeda database was used \citep{patu03}. When there were no measurements available, $\sigma=281$ km s$^{-1}$ was adopted \citep{birz04}.\\
$^d$Projected semimajor axis of the bubble.\\
$^e$Projected semiminor axis of the bubble.\\
$^f$Projected radial distance from the cluster core to the bubble's center.\\
\end{minipage}
\end{table*}
 
\subsection{Radio Properties}
Since the bubbles are thought to be the result of the interaction between the central radio source and the ICM, it is of interest to determine whether a central radio source is present. \citet{mitt09} performed an radio analysis for the HIFLUGCS sample using literature information (e.g., NVSS, VLSS, FIRST etc) plus archived VLA radio data. As a result, for the HIFLUGCS sample the presence of a central radio source is based on the \citet{mitt09} analysis except for A754, A1060, A1650 and A1736. These sources show the presence of a central radio source coincident with the BCG location (see Table \ref{radio_info_table} for references). \citet{mitt09} described a radio source as central if there is a radio source within $50 h_{71}^{-1}$~kpc of the X-ray peak, except A3562, A2142, A4038 and A3376 where a cut of $12 h_{71}^{-1}$~kpc was used. For the systems that are in B55 but not in the HIFLUGCS sample we established the presence of a central radio source based on information from the literature. We define a radio source as central if its location is coincident with the BCG location. Table \ref{radio_info_table} lists the presence of a radio source for our sample and 
the systems marked with an asterisk are the ones with central radio lobes. 
 
Also, Table \ref{radio_info_table} lists the 1.4 GHz monochromatic radio luminosities for the central radio sources, calculated as follows:
\begin{equation}
L_{1.4\rm{GHz}}=4 \pi D_{\rm{L}}^{2} S_{1.4\rm{GHz}} (1+z)^{\alpha-1},
\end{equation}
where $\alpha$  is the spectral index assuming $S_{\nu} \sim \nu^{-\alpha}$ and $S_{1.4\rm{GHz}}$ is the flux density at 1.4 GHz. \citet{mitt09} found that there is a good correlation between the total radio luminosity and the monochromatic radio luminosity (except for 2A 0335+096, A3376, MKW 3S and A4038).
 When available, the radio luminosities are from \citet{mitt09}. Otherwise, they are computed using information from the literature \citep[e.g. FIRST, TGSS, NVSS surveys,][]{VirL04,birz04,lind85,govo09,baum91,vent01}. In cases where a spectral index was not available, a value of -1 was adopted. For the systems with no detected central radio source, Table \ref{radio_info_table} lists the upper limits from the NVSS images (the numbers without errors in the above table) or from \citet{mitt09} for the HIFLUGCS sample.
 
\subsection{Merging Activity\label{merging}}
 
 Our systems range from  giant ellipticals and poor clusters to massive clusters, and they often reside in active environments with various degrees of merging activity. Since merging activity is part of cluster formation and evolution and might be different for cooling flow versus non-cooling flow systems, we searched the literature for such information for all our systems. This information is listed in Table \ref{radio_info_table}. However, we do not make any distinction between major mergers and minor ones. 
 
Signatures of merging activity can be seen in optical, X-ray, and radio data. In optical data, merging may be inferred from the detection of substructures in the cluster, such as local peaks in the spatial distribution of galaxies or localized velocity substructure. This is the case of A2142 \citep{ower11}, A1367 \citep{cort04}, A1689 \citep{leon11}, A2063 and MKW3S \citep{krem99}, A3395s \citep{donn01}, among many others (see Table \ref{radio_info_table}). 
 
Furthermore, the X-ray data can reveal shocks that might be related to merger activity \citep[e.g., A754,][]{henr04}, the presence of a significant velocity gradient \citep[e.g., A576,][]{dupk07}, detection of a low entropy gas \citep[e.g., A3266,][]{fino06}, and detection of cold fronts \citep{ower09,ghiz10} and sloshing activity \citep[e.g., A1795,][]{mark01}.  Cold fronts were first detected in \emph{Chandra} data \citep{mark00,vikh01} as contact discontinuities at the boundary of a cool, dense gas that is moving subsonically through a hotter, more diffuse surrounding gas \citep[for a review, see][]{mark07}. Using the B55 sample, \citet{ghiz10} found that a large fraction of systems have cold fronts. 

Sloshing was introduced by \citet{mark01} to explain cold fronts in the relaxed cluster A1795. Other examples of sloshing are found in A1644 \citep{john10} and Ophiuchus \citep{mill10}, among many others \citep{mark07}. 
Sloshing activity is thought to be a gravitational phenomenon, due to oscillations in the dark matter potential which might be set off by minor mergers \citep{asca06}, or hydrodynamical in nature \citep{chur03,fuji04}. Furthermore, \citet{zuho10} performed simulations of sloshing activity, and they show that such activity might be responsible for quenching the cooling in some cooling flow clusters.

Lastly, radio haloes and relics are indicators of merging activity. Haloes and relics are thought to be formed in major mergers \citep{weer11}, and radio mini-haloes may be indicators of sloshing \citep{zuho11b}. In Table \ref{radio_info_table}, systems which have a radio mini-halo, radio halo or a radio relic are indicated.
 
\section{X-ray Data Analysis}\label{analysis}
\subsection{General ICM Properties}\label{analysis_gen}

All systems were observed with the \textit{Chandra} ACIS detector in imaging mode, and the data were obtained from the \textit{Chandra} Data Archive. Details of the observations are given in Table \ref{tcool_table}.

The Chandra data were reprocessed with CIAO 4.2 using CALDB 4.2.1 and were corrected for known time-dependent gain and charge transfer inefficiency.  Blank-sky background files, normalized to the count rate of the source image in the $10-12$ keV band, were used for background subtraction.\footnote{See \url{http://asc.harvard.edu/contrib/maxim/acisbg/}.} 

The analysis of the X-ray data closely followed that of \citet{raff08}. Briefly, X-ray spectra were extracted in circular annuli with $\sim 3000$ counts centered on the centroid of the cluster emission.  For the fainter systems we used fewer counts (2000) for the inner region to allow derivation of the properties as close to the core as possible (e.g., Triangulum Aus, A2657, ZwCl 1215, III Zw 54, A2147, A1736, EXO 0423.4-0840, MKW 8). Also, in the case of the corona systems, due to the very small cores, we used for the inner region fewer than 3000 counts (2000 counts for A3558, 1000 counts for A3376, and 500 counts for A3395s, 3C129.1, Coma). In some  systems, due to the diffuse nature of the cluster emission or to substructure, the centroid was difficult to identify precisely. These systems are noted in Table \ref{info_table}. Spectra and their associated weighted responses were made for each annulus using CIAO and were fit in XSPEC version 12.5.1 \citep{arna96}.

Gas temperatures and densities were found by deprojecting the spectra with a single-temperature plasma model (MEKAL) with a foreground absorption model (WABS) using the PROJCT mixing model. We derived the cooling times using the deprojected densities and temperatures found above and the cooling curves from APEC spectral model \citet{smit01}. The pressure in each annulus was calculated as $p=nkT,$ where we have assumed an ideal gas and $n \approx 2n_{\rm{e}}$. Lastly, the entropy is defined as $S=kTn_{\rm e}^{-2/3}$ and has units of keV cm$^2$. 

To investigate whether or not the bubbles have enough energy to balance cooling, ones needs the luminosity of the cooling gas inside the cooling radius. Within the cooling radius, radiative energy losses must be replaced to prevent the cooling of large quantities of gas. We define the cooling radius as the radius  within which the gas has a cooling time less than 7.7 $\times$ 10$^{9}$ yr, the look-back time to $z=1$ for our adopted cosmology. To find the total luminosity inside the cooling radius, we performed the deprojection using a single-temperature model. Table \ref{tcool_table} gives values of $t_{cool}$ for all clusters, and  $r_{cool}$ and $L_{x}(<r_{cool})$  for only  our subsample of clusters that require heating (see Section \ref{Subsample}). 

To derive densities as close to the core as possible, we additionally use the onion peeling deprojection method described in \citet{raff08} inside the innermost annulus used for spectral deprojection, assuming that the temperature and abundance of the gas are constant in the inner region used in spectral deprojection and equal to the derived emission-weighted values. For a number of objects, constraints on the density profile inside the innermost annulus were not improved by this method (see Table \ref{tcool_table}). For these cases, we report only the emission-weighted values from the innermost annulus.

\subsection{Cooling Time, Temperature Drop and the Thermal Stability Criterion}\label{tcool_Tdrop_instab}

Our goal is to understand the biases and selection effects in the detectability of current X-ray bubble samples and to place limits on the fraction of systems that can have enough bubble power to balance cooling. As a result, from these complete samples we need a subsample of systems in which feedback is expected to operate (i.e., cooling flow systems). The cooling time in the core is a basic selection criterion for cooling flows: if the cooling time of a system is smaller than its age, which is typically a significant fraction of the Hubble time, then one expects that this system needs heating to prevent cooling. 

In order to calculate the central cooling times we used archival \emph{Chandra} data. The minimum radius at which we could derive reliable cooling times depends on the integration time and proximity of the source (see Table \ref{tcool_table} for the radius of the inner annulus used in the deprojection). Since we want to compare the cooling times for all the systems at a single physical radius, as close as possible to the nucleus, we computed the cooling time at 1 kpc using the surface brightness profiles to obtain densities (see Section \ref{analysis_gen}). For some systems the extrapolation did not work well, as the surface brightness profile is too noisy or drops towards the centre or there is significant substructure that is inconsistent with the deprojection method. This was the case for: A401, A1367, A1736, A2256, A3667, A2319, A119, ZwCl 1215, A2657, A3562, A644 and A2255. For some other systems, deprojection did not work because the surface brightness profile drops or flattens towards the centre or because they have X-ray-bright  AGNs (such as Perseus) or large bubbles. This is the case for A133, A478, A2597, Perseus, A2199, A4059 and Cygnus A. In these cases, the cooling time at 1 kpc was extrapolated from the cooling time profile from the single-temperature model deprojection. This extrapolation was generally small, since for all these systems the inner bin has a radius below 10 kpc. For some systems there is a large difference between the cooling time from the innermost annulus (used in spectral deprojection) and the cooling time at a radius of 1 kpc. In some cases this is due to a very peaked surface brightness profile. This is the case for MKW 8, A2634, A400, III Zw 54, A576, A1060, A3391, A2142, A3532, Coma, A3376, A3395s, and A3558. 
  
Additionally, we calculated the central temperature drop, since for a cooling flow cluster the temperature is expected to drop towards the center. We calculated the drop as the ratio between the highest temperature in the profile and the temperature of the innermost annulus. Table \ref{tcool_table} lists the temperature drop values for all systems. Some systems have no entry since the temperature profile increases towards the center, or the profile was too noisy and the temperature drop value was insignificant within errors. The calculated temperature drop depends on the size of the innermost annulus. However, the temperature typically varies slowly with radius, so it will not affect our estimates. Based on the temperature drop criterion only 2 systems with short cooling times at 1 kpc (less than 1 $\times  10^{9}$ yr) do not show a temperature drop (A1651 and A2063). However, the temperature declines are probably related to the local virial temperature, so the cooler clusters, where the virial temperature of the central galaxy is closer to that of the cluster, have smaller central temperature declines. The temperature decline is not used in this paper as a criterion to separate the cooling flow systems from non-cooling flow systems, but just as an additional test after the systems where already separated by using a different criterion, such as cooling time.  

An alternative way to select cooling flows is based on the thermal stability of the gas \citep{voit08,shar12}. Using the sample of clusters, \citet{voit08} found that star formation and H-$\alpha$ emission (and hence cooling) seem to occur only if, at some location in the cluster, the following condition is met:
\begin{equation}
{\eta_{\rm min}=\rm min}\left(\frac{\kappa T}{\Lambda (T) n_{e} n_{H} r^{2}}\right) \sim \frac{1}{f_{c}} \lesssim 5,
\end{equation}
where $\Lambda(T)$ is the cooling function calculated using APEC spectral model \citep{smit01},  and $f_{c}$ is the factor by which the magnetic field suppresses the conductivity below the Spitzer value. Assuming that the effective thermal conductivity can be expressed as a multiple, $f_{c}$, of the Spitzer value, this parameter provides a measure of the  stability of the gas to local cooling. For large values of $\eta_{\rm min}$, thermal conduction overwhelms radiative cooling, preventing local cooling throughout the ICM. For small values of this parameter, local cooling can run away, so that parts of the ICM may cool to low temperatures, resulting in the deposition of the cooled gas. If AGN are powered by the accretion of hot gas, then  $\eta_{\rm min}$ should have little significance for the feedback process. However, if the AGN are fueled by cooled ICM and the effective conductivity is comparable to the Spitzer value, $\eta_{\rm min}$ determines the systems where fuel may be available. 

\citet{voit08} found that values of $\eta_{\rm min} \lesssim 5$ correspond approximately to an inner cooling time of $5\times 10^{8}$ yr. This agrees well with our findings (see Table \ref{tcool_table} for the $\eta_{\rm min}$ values), the exeptions are A2065, A3391 and A400 which have  $\eta_{\rm min} < 5$, but central cooling times of  $\gtrsim 5 \times 10^{8}$ yr. However the last 2 systems are corona systems \citep[with small cool cores, e.g.,][]{sun07} with shallow \emph{Chandra} observations and we were not able to get closer to the mini-core.

\section{Results from X-ray and Other Observations}\label{S:results_xray}

In this section, we examine how the central cooling time and $\eta_{\rm min}$ relate to other properties of the cluster, such as the separation between the X-ray centroid and the optical centre (an indication of how relaxed a system is) and the central radio source luminosity. 

\subsection{Cooling Time and Radio Luminosity\label{disc:tcool_lrad}}

 Figure \ref{F:tcool_instab_vs_lrad} left  shows the central cooling time at 1 kpc versus the monochromatic 1.4 GHz luminosity of the central radio source. There is a separation between clusters, such that objects with radio luminosities above 2.5 $\times$ 10$^{30}$ erg s$^{-1}$ Hz$^{-1}$ generally have cooling times at 1 kpc less than $ 5 \times 10^{8}$ yr and have bubbles. There are no high radio luminosity objects with longer cooling times. A similar plot is showed in  \citet{cava08}, but they were no able to probe the CF/NCF separation threshold since they used NVSS radio fluxes, and were not able to distinguish between central radio source emission and halo emission. \citet{sun09} found a similar cutoff in radio luminosity for their strong cool-core systems (with cooling times below 1 Gyr). 

\citet{raff08} studied star formation in a sample of 39 systems in different cooling stages and found a sharp threshold for the onset of star formation that occurs when the central cooling time falls below $\sim 5 \times 10^{8}$ yr \citep[see also][]{cava08}.  This number agrees well with our results shown in Figure \ref{F:tcool_instab_vs_lrad} left. The only objects in our sample with a central cooling time $\gtrsim 5 \times 10^{8}$ yr and with a high radio luminosity ($\gtrsim 2.5 \times$ 10$^{30}$ erg s$^{-1}$ Hz$^{-1}$) are A3391, A400 and A1689. The first two are corona systems for which we were not able to probe the cooling time in the tiny cores. For A3391 we see bubbles in the existing \emph{Chandra} image. The fact that there are no objects with high radio luminosity ($ > 2.5 \times$ 10$^{30}$ erg s$^{-1}$ Hz$^{-1}$) and long cooling times ($> 5 \times 10^{8}$ yr) indicates that the central AGN radio-mode heating in clusters is sensitive to the presence of gas with short cooling times. This finding fits well with the AGN feedback scenario.

Also in Figure \ref{F:tcool_instab_vs_lrad} we plotted in red the systems that have evidence in the literature for merger activity. We do not differentiate between a minor or major merger or sloshing (see Section \ref{merging} for details). Figure \ref{F:tcool_instab_vs_lrad} shows that mergers are present in both cooling-flow and non-cooling flow systems.  We note, however, that the typical strength of the merger likely differs between the two classes.

\subsection{Thermal Stability Parameter and Radio Luminosity\label{disc:instab_lrad}}  

\begin{figure*}
\begin{tabular}{@{}cc}
\includegraphics[width=84mm]{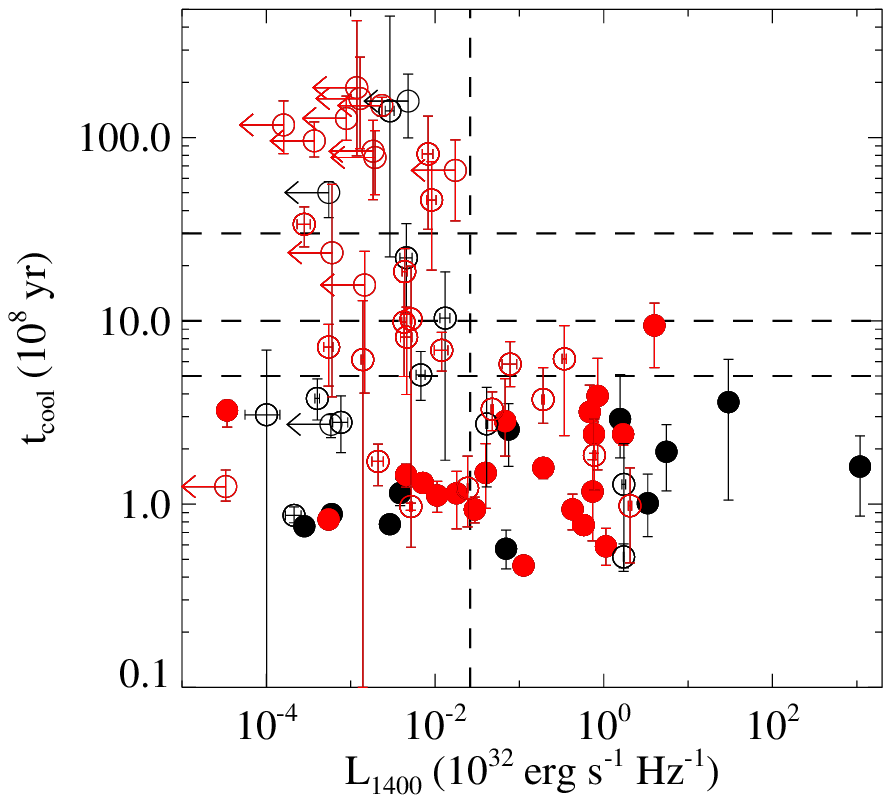} &
\includegraphics[width=84mm]{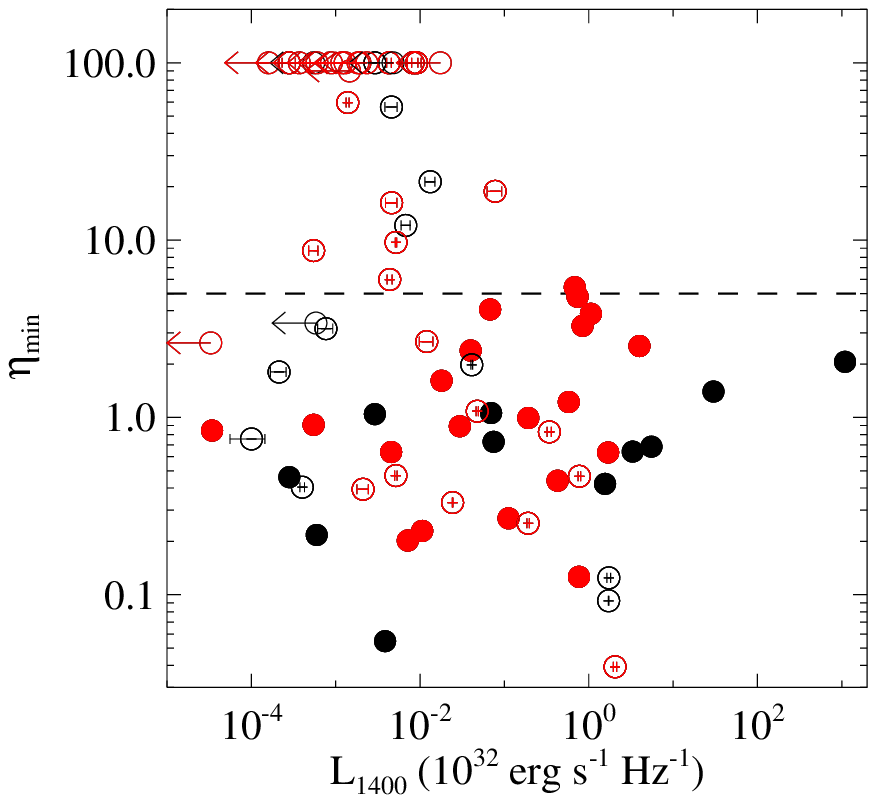} \\
\end{tabular}
\caption{\emph{Left}: Cooling time (at 1~kpc) versus the monochromatic 1.4 GHz radio luminosity for all 75 systems from both B55 and HIFLUGCS. The radio luminosity is taken from \citet{mitt09} were available, otherwise was computed  using fluxes from the literature (see Table \ref{info_table} for details and references). The filled symbols denote the 31 systems with bubbles (see Section \ref{Cav_sect}). The red symbols denote the systems that appear to be undergoing some merger activity such as major mergers, minor mergers, or sloshing. This information is taken from the literature (see Table \ref{tcool_table} for references). Over-plotted are the lines at $5 \times 10^{8}$ yr, $1 \times 10^{9}$ yr, $3 \times 10^{9}$ yr and 2.5 $\times$ 10$^{30}$ erg s$^{-1}$ Hz$^{-1}$ (see text for details). \emph{right}: Thermal stability parameter (see Section \ref{tcool_Tdrop_instab}) versus the monochromatic 1.4 GHz radio luminosity. The symbols are the same as on the left panel.}
\label{F:tcool_instab_vs_lrad}
\end{figure*}

\citet{voit08} argued that a central cooling time of $\sim 5 \times 10^{8}$ yr  is equivalent to a value of 5 for $\eta_{\rm min}$, which agrees well with our results, shown in Figure \ref{F:tcool_instab_vs_lrad} right, which plots $\eta_{\rm min}$ versus the radio luminosity. A3391, A400 and A2065 have central cooling times higher than $\sim 5 \times 10^{8}$ yr and $\eta_{\rm min}$ smaller than 5. However the first two systems, as mentioned in the discussion of Figure \ref{F:tcool_instab_vs_lrad} left, are corona systems with shallow \emph{Chandra} observations and we were not able to get closer to the mini-core. On the other hand, A1689 has a central cooling time of $\sim 5 \times 10^{8}$ yr, but $\eta_{\rm min}$ much larger than 5.

\subsection{Cooling Time and X-ray Peak/Optical Core Separation}\label{disc_sep}

\begin{figure}
\includegraphics[width=84mm]{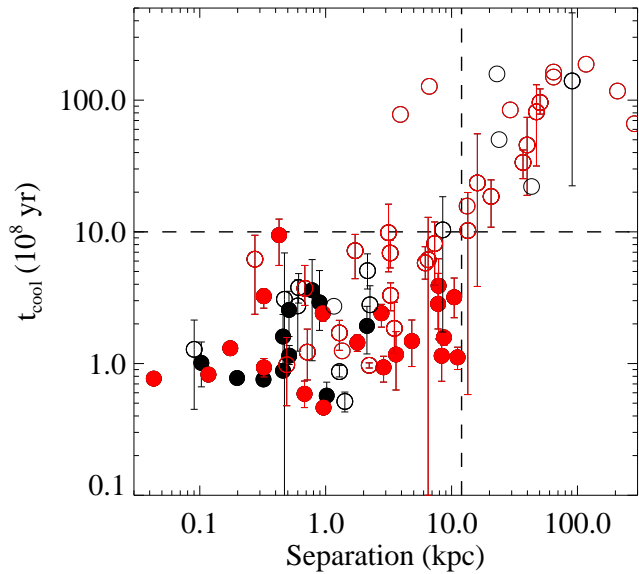}
\caption{Central cooling time (at 1 kpc) versus the separation between the X-ray peak and the BCG center. Filled symbols denote the systems with bubbles and red colors denote the merging systems. Also plotted are the lines at $5 \times 10^{8}$ yr and 12 kpc.}
\label{F:sep_tcool}
\end{figure}

Figure \ref{F:sep_tcool} shows a strong relationship between central cooling time and the separation between X-ray centroid and the optical centre (see Table \ref{info_table}). No object with central cooling time less than $\sim  10^{9}$ yr has a separation above 12 kpc, and almost no objects with a separation above 12 kpc (exceptions A401, A119) have a cooling time below $\sim 10^{9}$ yr. For the systems with large separation between the optical centre and the X-ray centroid indicates that these systems are not dynamically settled that they are still going through significant merger activity. This trend is similar to the one in Figure \ref{F:tcool_instab_vs_lrad} left, but  much tighter. The only system with a cooling time below $\sim  10^{9}$ yr that has a separation higher than 12 kpc is A4038. This system might have experienced major merger activity, which might have been created the radio relic present on the outskirts of the cluster \citep{slee98}. Furthermore, \citet{raff08} shows that the systems with larger separation do not have star formation. This is another indicator that these systems do not need heating.

\section{The AGN Heating Duty Cycle}
\subsection{Subsample of Clusters that Require Heating}\label{Subsample}
Our goal is to understand the biases and selection effects in the detection of X-ray bubbles, and to place limits on the fraction of systems that can have enough bubble power to balance cooling (these are the cooling-flow systems, CFs). However, in systems in which no cooling is expected to occur (i.e., non-cooling flows, NCFs), one also expects no heating. Therefore, these systems should not be considered in our analysis. As a result, a subsample in which feedback is expected to operate must be defined from these complete samples. 

In Figure \ref{F:tcool_instab_vs_lrad} (left panel) we found a clear separation between CF and NCF systems. Furthermore, we found that the criteria using the thermal stability parameter and cooling time to separate CF and NCF systems give similar results. Based on these results, we now define a subsample requiring heating based on the thermal stability parameter ($\eta_{\rm min}$ less than 5), since this criterium is more physically motivated. Additionally, based on the clear separation between clusters seen in Figure \ref{F:sep_tcool} with a cut-off value of inner cooling time (at 1 kpc) of $1 \times 10^{9}$ yr, and a cut-off value of 12 kpc for the projected separation between the X-ray peak and optical centroids, we further limited the subsample to systems that have such a separation of less than 12 kpc. In total, there are 49 systems (32 for B55 and 41 for HIFLUGCS) that meet these criteria ($\eta_{\rm min}< 5$ and separation $<$ 12 kpc), 31 of which have detected bubbles (22 for B55, 26 for HIFLUGCS). Table \ref{sim_table} lists the systems that meet these criteria and also lack detected bubbles.

Finally, based on the central cooling time (at 1 kpc) and the separation (see Figure \ref{F:sep_tcool}), we also consider a somewhat larger subsample: those systems with central cooling times below $\sim 10^{9}$ yr and separations smaller than 12 kpc. This subsample includes the previous main heating subsample, plus 7 extra systems (see Table \ref{sim_table}). There are a total of 56 systems (39 for B55, 47 for HIFLUGCS) in this extended subsample. Therefore, there are 16 systems in B55 and 16 systems in HIFLUGCS (19 systems in total)  that are excluded from the samples of systems that need heating.

\subsection{Simulations}\label{sim}
For the systems without bubbles from our extended subsample (which have cooling times at 1 kpc smaller than $\sim 10^{9}$ yr), we performed simulations in order to find the locations and the sizes of bubbles that can be present in the cluster and remain undetected. These systems are listed in Table \ref{sim_table}.

No attempt was made to simulate the effects of
bright rims around the bubbles, since that would complicate the model
greatly.  When present, these can enhance bubble contrast, making them
more evident.  As a result, if undetected bubbles had bright rims, we
may be overestimating their possible energy content.  Regardless of
this, the modeling used here provides good limits on the energy that
may be contained in undetected bubbles.

We simulated a 3 dimensional cluster using a spherically symmetric $\beta$ profile model for the emissivity (derived from the existing archive observations, see Table \ref{sim_table}). The parameters of the $\beta$ model were fixed to those derived from the azimuthally averaged surface brightness profiles. When a double-$\beta$ model was preferred over a single-$\beta$ model, the double-$\beta$ model was used in the simulation. A cube of $500\times 500\times 500$ elements was populated with the corresponding emissivities. 

For simplicity, spherical bubbles were simulated in this cube by setting the emissivities inside the bubbles to zero. However, it is important to notice that observationally it is unclear whether the X-ray cavities are devoid of X-ray emitting gas or instead are filled with hot, underdense gas. Then, by integrating the emissivity along lines of sight through the cube, we obtained the 2 dimensional surface brightness map. This map was used as input for the MARX simulator\footnote{see \url{http://www.space.mit.edu/CXC/MARX/}} in order to obtain a simulated \emph{Chandra} image. For MARX we used a spectrum extracted from the data, with the integration time and position set as in the data. As in the case of the systems from Section \ref{Cav_sect}, we used the \emph{ciao} tool \emph{merge\_all} to add all the available observations for each system. This total integration time is used in the simulations (see Table \ref{sim_table} for details).

We assume that the bubbles expand adiabatically into a $\beta$-model atmosphere. Therefore, the radius of the bubble at at the distance $r$ from the centre of the cluster is given by:
\begin{equation}
R_{\rm b}(r)=R_{\rm b}(0) \left[1+\left(\frac{r}{r_{\rm c}}\right)^2 \right]^{3 \beta/10},
\end{equation}
where $R_{\rm b}(0)$ is the bubble size at the cluster center, $r_{\rm c}$ is the core radius and $\beta$ is the parameter of the beta model.
To calculate the ages  we assume that the bubbles rise buoyantly from the centre of the cluster to their current radius (giving $t_{\rm{buoy}}$, see section \ref{Cav_sect} for details). In order to calculate the initial size, $R_{\rm b}(0)$, with which the bubbles are injected, we assumed: $4pV \sim P_{\rm{cav}}t \sim L_{\rm{X}}t$, where $p$ is the central pressure of the cluster,  $L_{\rm{X}}$ is the bolometric X-ray luminosity inside $r_{\rm cool}$ (see Section \ref{analysis}) and $t$ is the time between the outbursts. We adopted a time between outbursts of $10^{8}$~yr, motivated by the observations of clusters with multiple generations of bubbles, such as Perseus \citep{fabi00}.  We ran multiple simulations for a cluster, with the locations of the bubbles and the angle in the plane of sky ($\phi$) randomized, and with an angle from the line of sight ($\theta$) of $90^\circ$. The resulting images were inspected to determine if the bubbles remains visible for the whole of the $10^8$~yr outburst cycle. If a bubble become undetectable when its buoyancy age was less than $10^8$~yr, such a bubble may be present, but undetectable, in the real cluster and the heating rate could be sufficient to balance cooling. In cases where the bubble would have remained visible for the whole cycle, the simulation was repeated using smaller values of $P_{\rm{cav}}$, until the bubble becomes undetectable in less than $10^{8}$~yr. Under our assumptions, this value of $P_{\rm{cav}}$ determines an upper limit on the possible AGN power in such systems.  These systems are A3112, A1650, A1689, RXCJ 1504.1-0248, A2244, and Ophiuchus. Therefore, these systems have upper limits less than the value required to balance cooling under our assumptions (see Table \ref{sim_table}).

\begin{table*}
\begin{minipage}{145mm}
\caption{Simulation Parameters}
\label{sim_table}
\begin{tabular}{@{}lcccccccc}
\hline
   & $t$$^b$ &  $\sigma$$^c$ &  $r_{\rm c1}$ & $\beta_{1}$ & $I_{\rm o}$$^d$ & $I_{\rm bk}$$^e$   &  $r_{\rm inj}$$^f$ & Result \\
System$^a$ &  (ks) &  (km s$^{-1}$)  &  (kpc) &    &    &     &  (kpc) &   \\
\hline
 \multicolumn{9}{c}{Main subsample: (Minimum instability $< $5) and (Separation $<$ 12 kpc)} \\
 \hline
 AWM 7           & 180  & 330       & 6.8          & 0.31        & 121        & 5.0  & 4.7  & $L_{\rm X}$ \\
 A400*           & 21.5 & 301      & 135.6        & 0.39        & 5.4        & 0.5  & 4.3  & $L_{\rm X}$ \\
 A3112           & 109  & \ldots  & 15.2         & 0.43        & 1091       & 1.0  & 10.5 & $L_{\rm X}/5$  \\
 EXO 0423.4-0840 & 25.9 & \ldots  & 10.1         & 0.44        & 428        & 0.17 & 6.4  & $L_{\rm X}$ \\
 3C129.1*        & 9.6  & \ldots   & 66.3         & 0.33        & 11.5       & 7.0  & 6.4  & $L_{\rm X}$ \\
 A3376*          & 60   & 268      & 136.1        & 0.80        & 5.7        & 1.3  & 5.7  & $L_{\rm X}$ \\
 A3395s*         & 19.8 & 335       & 67.5         & 0.38        & 8.9        & 1.4  & 3.8  & $L_{\rm X}$ \\
 MKW 4           & 30   & 275       & 2.5          & 0.43        & 455        & 3.1  & 4.1  & $L_{\rm X}$ \\
 A1644           & 70   & \ldots   & 3.6          & 0.31        & 160        & 1.0  & 7.4  & $L_{\rm X}$ \\
 A1650           & 228  & \ldots   & 35.5         & 0.42        & 188        & 0.41 & 13.0 & $L_{\rm X}/5$ \\
 Coma (NGC4874)* & 9.5  & 278       & 140.7        & 0.38        & 35.8       & 10.0 & 6.0  & $L_{\rm X}$ \\
 A3558*          & 11.2 & 258       & 37.5         & 0.34        & 72.3       & 5.0  & 11.0 & $L_{\rm X}$ \\
 MKW 8*          & 23.1 & 217       & 53.4         & 0.31        & 8.1        & 0.5  &  3.1 & $L_{\rm X}$ \\
 RXCJ 1504.1-0248& 53   & \ldots  & 36.7 (145.8) & 0.75 (0.74) & 3319 (142) & 1.5  & 18.4 & $L_{\rm X}/10$ \\
 A2065           & 50   & \ldots   & 53.0         & 0.41        & 65.4       & 0.5  & 10   & $L_{\rm X}$ \\
 Ophiuchus       & 51   & \ldots   & 4.3          & 0.27        & 416        & 5.0  & 12.0 & $L_{\rm X}/5$ \\
 A2589*           & 93   & 279       & 3.0 (751.5)  & 0.29 (21)   & 108 (13.6) & 2.5  & 8.5  & $L_{\rm X}$ \\
 A2634 (3C465)*  & 49.5 & 359      & 0.05         & 0.29        & 397        & 4.2  & 4.1  & $L_{\rm X}$ \\
 \hline
\multicolumn{9}{c}{Extra systems: (Cooling time at 1 kpc: 0.5-1 Gyr) and (Separation $<$ 12 kpc)} \\
\hline
 A576*           & 25.6 & \ldots   & 20.0         & 0.30        & 34.7       & 1.0  & 6.6  & $L_{\rm X}$ \\
 A1651           & 8.9  & \ldots   & 82.7         & 0.51        & 103        & 0.29 & 9.8  & $L_{\rm X}$ \\
 A1689           & 194  & \ldots  & 61.0         & 0.54        & 420        & 0.54 & 16.0 & $L_{\rm X}/5$  \\
 A3571           & 31  & 323        & 10.9 (47.3)  & 0.54 (0.33) & 77.7 (91)  & 5.0  & 5.6  & $L_{\rm X}$ \\ 
 A2063           & 50  & 222        & 21.2         & 0.35        & 73         & 1.6  & 8.4  & $L_{\rm X}$ \\
 A2142           & 44.6& \ldots   & 67.4         & 0.48        & 27         & 1.0  & 18.2 & $L_{\rm X}$ \\
 A2244           & 56.7& \ldots   & 74.3         & 0.55        & 147        & 2.3  & 12.6 & $L_{\rm X}/5$ \\
\hline
\end{tabular}\\
$^a$Systems without bubbles in the existing \emph{Chandra} images, but with a separation $<$ 12 kpc and $\eta_{\rm min} < 5$ or a central cooling time (at 1kpc) $<$ 1 Gyr. The asterisks denote the corona class systems \citep{sun07,sun09}.\\
$^b$The total integration time from multiple observations when available: AWM 7 (Obs IDs 908, 11717, 12017, 12018), A3112 (Obs IDs 2516, 2216, 6972, 7324, 7323), EXO 0423.4-0840 (Obs IDs 3970, 4183), A1644 (Obs IDs 2206, 7922), A1650 (Obs IDs 4178, 5822, 5823, 6356, 6357, 6358, 7242), A1689 (Obs IDs 540, 5004, 1663), RXCJ 1504.1-0248 (Obs IDs 4935, 5793), A2063 (Obs IDs 4187, 5795, 6262, 6263) and A2589 (Obs IDs 7190, 3210, 6948, 7340).  Otherwise, we listed the time on source after re-processing  (the same as the ones in Table \ref{tcool_table}).\\
$^c$Velocity dispersions were taken from the HyperLeda database \citep{patu03}. When none were available, $\sigma=281$ km s$^{-1}$ was adopted  \citep{birz04}.\\
$^d$The normalisation of the isothermal surface brightness profile or $\beta$-profile \citep{cava76}  in units of 10$^{-9}$ ph s$^{-1}$ cm$^{-2}$ pix$^{-2}$.\\
$^e$The background contribution for the isothermal surface brightness profile in units of 10$^{-9}$ ph s$^{-1}$ cm$^{-2}$ pix$^{-2}$.\\
$^f$This is the bubble injection radius for the first run ($L_{\rm x}$, see Table \ref{tcool_table}); the injection radius for the actual run is 5.8 kpc for RXCJ 1504.1-0248 ($L_{\rm x}/10$), 5.8 kpc for A1650 ($L_{\rm x}/5$), 4.7 for A3112 ($L_{\rm x}/5$), 7.1 for A1689 ($L_{\rm x}/5$), 5.4 for Ophiuchus ($L_{\rm x}/5$), 5.64 kpc for A2244 ($L_{\rm x}/2$).\\
\end{minipage}
\end{table*}
\subsection{Biases in Bubble Samples}
Samples of observed bubbles could suffer from biases due to detectability and projection effects. To assess the degree to which such biases might be present, we simulated an A4059-like cluster with two bubbles with a variety of sizes and positions. We then attempted to detect the bubbles by eye (without knowing their locations or sizes a priori).  The measured properties can then be compared to the intrinsic ones to understand possible biases. 

\begin{figure*}
\begin{tabular}{@{}cc}
\includegraphics[width=84mm]{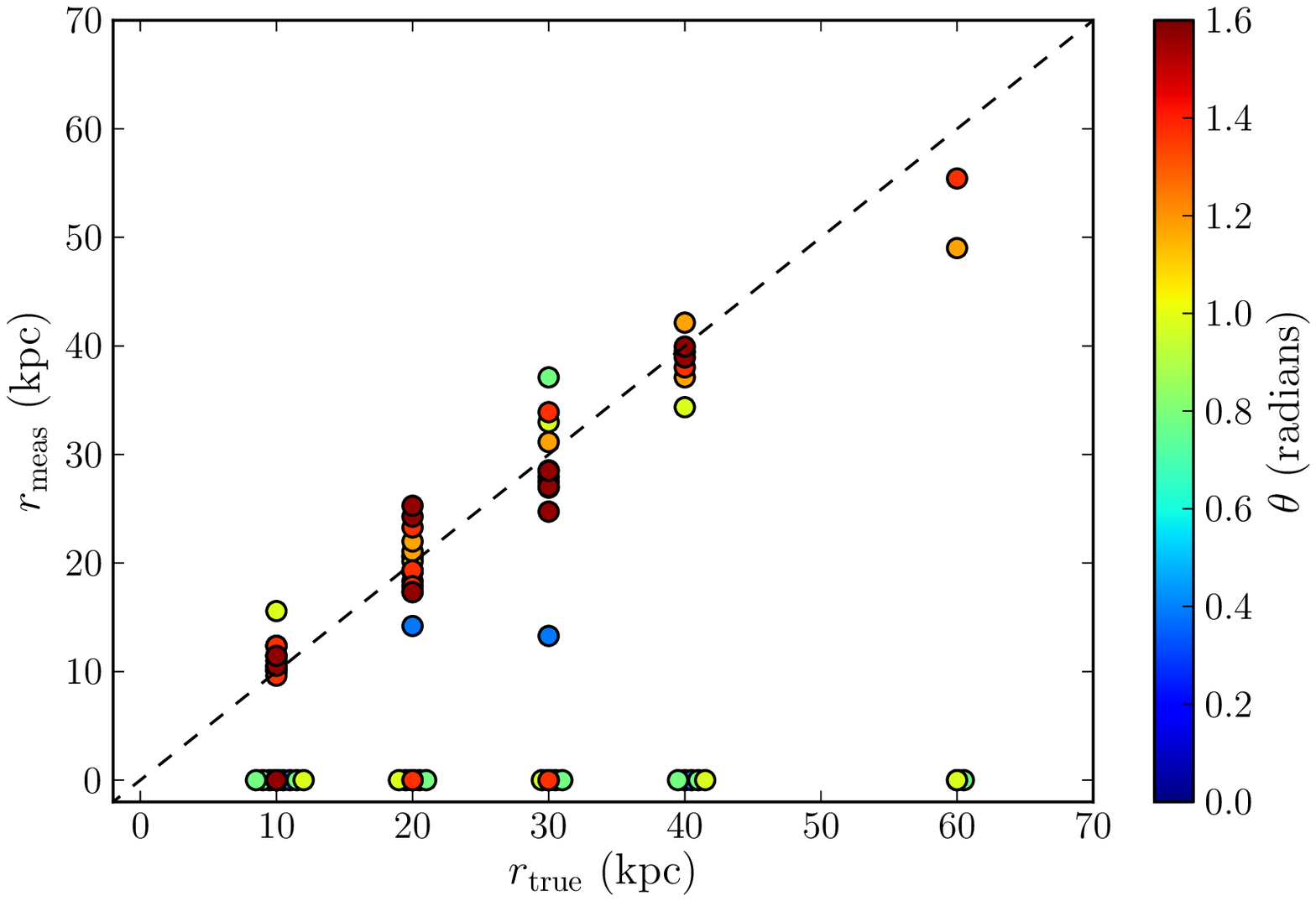} &
\includegraphics[width=84mm]{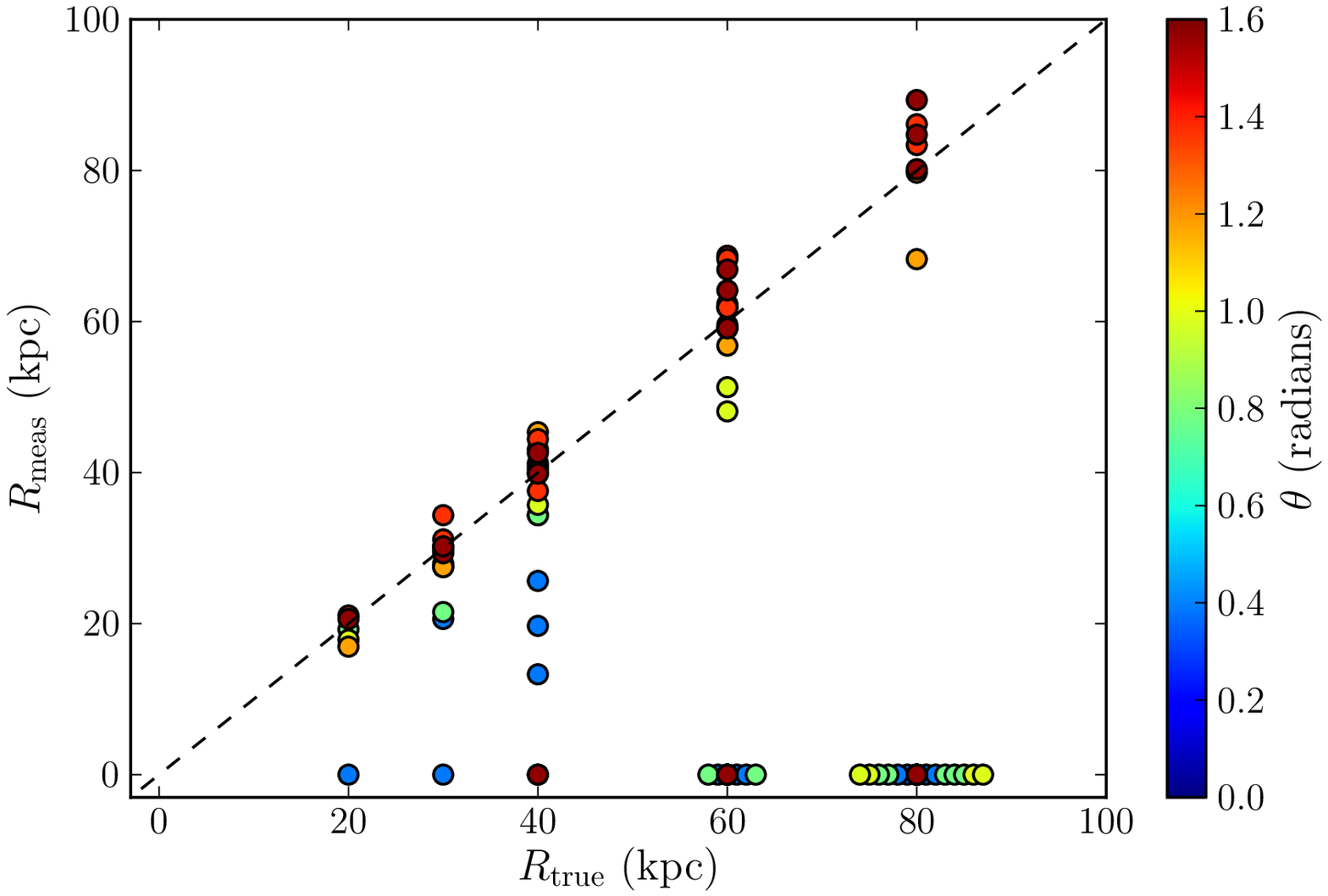} \\
\end{tabular}
\caption{\emph{Left:} Measured bubble radius versus the true radius. The color denotes the value of $\theta$, the angle between the bubble-to-core axis and the line of sight. Undetected bubbles are shown at a measured radius of zero. \emph{Right:} Measured distance from the core to the bubble versus the true distance.}
\label{F:A4059_sim}
\end{figure*}

In total, 89 simulations were performed. Bubble radii were varied between 10 kpc and 60 kpc, and the distance from the core to the bubble was varied between 20 kpc and 80 kpc. Bubbles with radii greater than the bubble-to-core distance were not simulated. Figure~\ref{F:A4059_sim} left shows a comparison between the measured bubble radii and the true ones. Approximately 44\% of the simulated bubbles were undetected. Of these, the majority had an axis between the bubble and core that was oriented near to the line of sight ($\theta \lesssim 45^{\circ}$) or were small bubbles at large distances from the core. Of those that were detected, the measured bubble radius is on average close to the true one. The greatest deviations from equality are generally for bubbles with smaller angles to the line of sight.

A similar situation also applies to the comparison between the measured distance to the core and the true one, shown in Figure~\ref{F:A4059_sim} right. It is clear that most of the undetected bubbles lie at large distances from the core or, again, have small values of $\theta$. For detected bubbles, the average distance is close to the true one in most cases. These results imply that when bubbles are detected, their properties are probably not grossly in error, particularly in an average sense. 

\begin{figure}
\includegraphics[width=84mm]{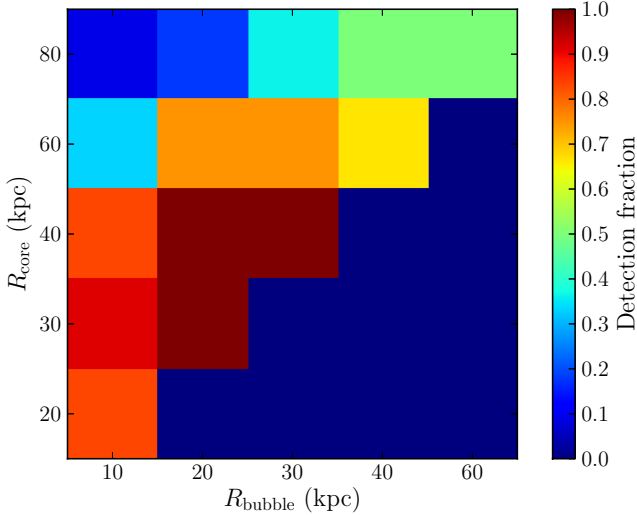}
\caption{Detection fraction as a function of the true bubble radius and the true (unprojected) distance from the bubble centre to the cluster core. The fraction has been summed over all angles between the bubble-to-core axis and the line of sight.}
\label{F:A4059_det}
\end{figure}

Lastly, we can estimate the detection rate for bubbles as a function of bubble radius and distance to the core by summing over all angles of the bubble-to-core axis to the line of sight. This detection rate is shown in Figure~\ref{F:A4059_det}. As expected, the detection fraction depends strongly on the bubble size and on the distance from the core. However, it is currently unclear as to which fraction of real bubbles are expected to fall into this parameter space.

\subsection{The Duty Cycle}\label{S:duty_cycle_results}
\begin{figure*}
\begin{tabular}{@{}cc}
\includegraphics[width=84mm]{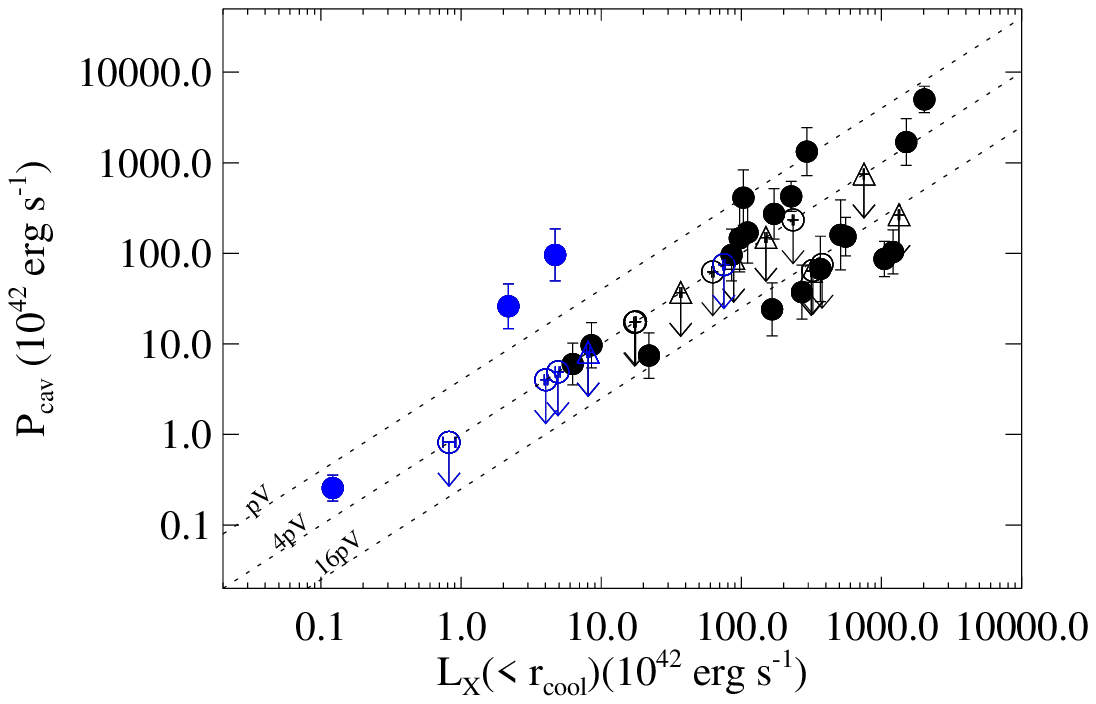} &
\includegraphics[width=84mm]{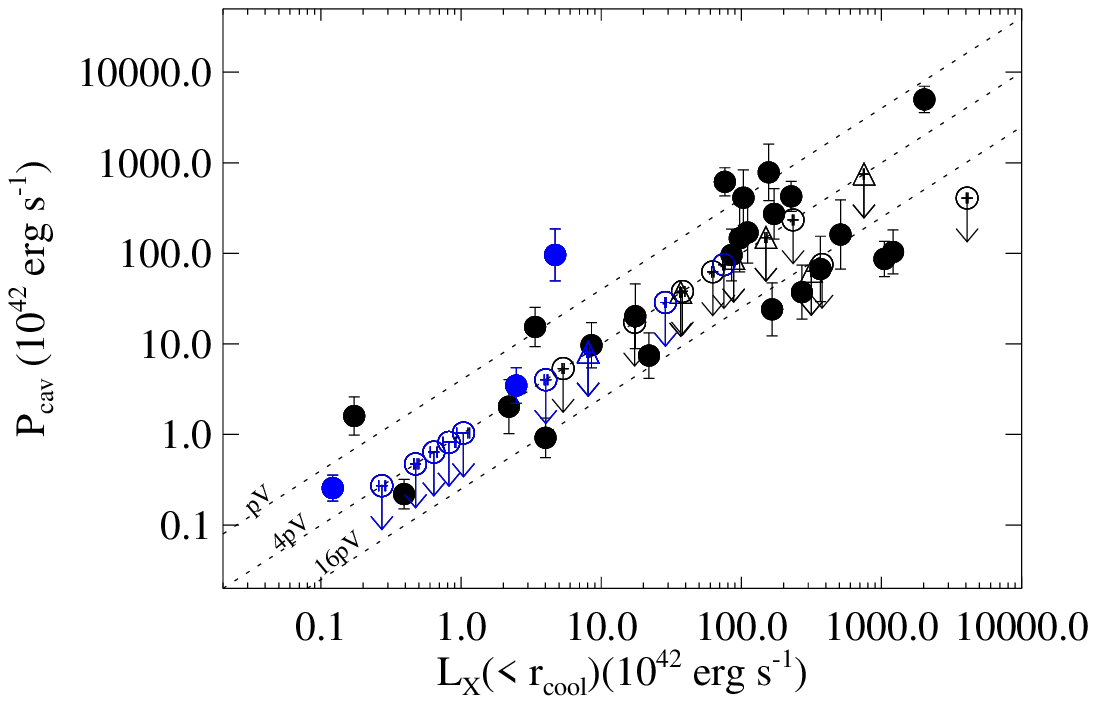} \\
\end{tabular}
\caption{Bubble power versus the X-ray luminosity (inside the cooling region) for the subsamples of systems that require heating from B55 (\emph{left} panel) and HIFLUGCS (\emph{right} panel). The subsample of systems that require heating is listed in Table \ref{sim_table} (see Section \ref{Subsample} for details). Different symbols denote different subsamples: circles for the main subsample ($\eta_{\rm min}<$ 5 and separation $<$ 12 kpc) and triangles for the extended sample (cooling time at 1 kpc between 0.5-1 Gyr and separation $<$ 12 kpc). The filled symbols denote the systems with bubbles, and the blue  symbols denote the corona-class systems, which are marked with an asterisk in Table \ref{sim_table}. The diagonal lines indicate $P_{\rm{cav}}=L_{\rm{X}}$ assuming $pV$, $4pV$ or $16pV$ as the energy deposited. The limits on the $4pV$ line are not strictly upper limits, but simply mark the systems where the limit is at least this large.}
\label{F:pcav_lx_B55_HIF}
\end{figure*}

Figure~\ref{F:pcav_lx_B55_HIF} shows the bubble power (the heating rate) versus the bolometric X-ray luminosity (the cooling rate) within the cooling radius (at which $t_{\rm{cool}}<7.7 \times  10^{9}$~yr) for our sample of cooling flows. The fraction of systems with detected bubbles for the main subsample  is $\approx$ 0.69 ($22/32$) for the B55 and $\approx$ 0.63 ($26/41$) for HIFLUGCS. We note that this sample is based on two criteria: $\eta_{\rm min}< 5$ and X-ray-to-optical core separation $<$ 12 kpc. If, however, the AGN are not fueled by cooled gas, then the $\eta_{\rm min}$ criterion is not relevant to this analysis. For the larger samples based on cooling time cut-off the fraction is $\approx$ 0.56 ($22/39$) for the B55 and $\approx$ 0.55 ($26/47$) for HIFLUGCS. This implies a duty cycle in cooling flows of at least 55\% to 69\%. This duty cycle agrees well with previous findings \citep{dunn06, nuls09, dong10} and with the radio-loud fraction of BCGs in cooling flows \citep{burn90}. It also agrees with the radio-loud fraction of $\sim 30$\% found for massive galaxies by \citet{best05}, when adjusted for the fraction of such systems that are in cooling flows ($\sim 50$\%, see Section~\ref{Subsample}).
  
From Figure~\ref{F:pcav_lx_B55_HIF} we can conclude that most of the systems that lack detected bubbles could have enough bubble power to balance cooling and remain undetected in existing images. A few systems have upper limits on the bubble power below the $4pV$ line, implying that they cannot have sufficient bubble power to balance cooling under our assumptions. However, one of our primary assumptions is that the bubbles are in the plane of the sky. As noted in Section \ref{intro}, the orientation of the bubble axis relative to the line of sight has a strong effect on the detectability of the bubbles. If the angle from the bubble axis to the line of sight is $\lesssim 30^{\circ}$, it is difficult to detect even large bubbles. As such an orientation is expected to occur in roughly 15\% of all systems (assuming the jet axes are randomly oriented), it is possible that $\sim 2$--3 of the systems without detected bubbles in our sample have significant bubble power that is hidden. Unfortunate orientation might therefore account for all of the systems for which the upper limit on the bubble power is insufficient to balance cooling. Therefore, with existing data, one cannot exclude the scenario that almost all cooling flow clusters have bubbles with enough power to balance cooling, implying a duty cycle for such activity of up to 100\%. 

However, of the 7 intermediate cooling flow systems (with a central cooling time between $\sim 5$ -- $10 \times 10^{8}$ yr), only A2063 shows extended radio emission. Thus, if the remaining intermediate systems do have bubbles, they must be  ``ghost'', like those of A2597 \citep{mcna01}. It is more probable that these systems have previously had a cooling core that was heated by a merger, such as A2142, A3571, A1651, A2244, and  corona system A576 (see Section \ref{disc_cf_vs_ncf}). Therefore, these systems might not need bubbles in order to balance the internal heating.

Also, it is important to mention that some of the systems without bubbles are corona-class systems (denoted by the blue symbols in Figure \ref{F:pcav_lx_B55_HIF}), which often have large radio lobes (all such systems in our sample except A576, MKW 8, A2589 and A3558 have lobes) embedded in lower density gas surrounding the bright X-ray core, which reduces the contrast of any bubbles and makes them harder to detect. Such systems may well have undetected bubbles. They include A400, A3376, A2634, A3395s, Coma, and 3C129.1. 

However, it is not fully understood what makes these systems different from larger cooling cores or whether the radio mode feedback cycle applies to them. Since the radio lobes are inflated beyond the cooling region and the fraction of the outburst energy coupled to the corona is small, their feedback process may be very inefficient, and heating required to balance the internal cooling might be supplied by SNe or by heat conduction \citep{sun07}. 
Alternatively, radio heating through weak shocks may be important in these systems \citep[for a discussion on the importance of weak shocks for AGN heating, see the review of][]{mcna12}. When the corona class systems are excluded, we find a duty cycle of 76\% ($=(22-3)/(32-7)$) for the B55 main subsample and 77\% ($=(26-3)/(41-11)$) for the HIFLUGCS main subsample.

Lastly, we find that the fraction of systems with detected bubbles for both complete samples is very similar. The two subsamples overlap for 44 systems (see Section \ref{comp_samples}), however the HIFLUGCS sample extends to lower-mass systems. In general, our results are most relevant for clusters, since there are only 10 ellipticals and poor clusters in our sample (3 for B55 and 9 for HIFLUGCS), and 8 of them have bubbles. 

\section{Discussion}
\subsection{Radio Luminosity Cut-Off}
Figure \ref{F:tcool_instab_vs_lrad} left shows that there is a radio luminosity cut-off for massive clusters, such that systems with a central radio luminosity above 2.5 $\times$ 10$^{30}$ erg s$^{-1}$ Hz$^{-1}$ are found only in cooling flows (see Section \ref{disc:tcool_lrad}). This supports the idea that the radio-mode outbursts are self-regulating, with increased radio luminosity (and hence feedback power) seen in sources where cooling is likely occurring. 

We also note that the luminosity of central radio sources does not appear to be lower on average when there are no detected bubbles. In such systems, the radio source lacks bright lobes and is core dominated but is generally as bright as the systems with radio lobes and bubbles. This is the case of A1689, A1644, RXCJ 1504.1-0248, which represent $ \sim 10 \%$ of the high-luminosity objects (see Section \ref{disc_cav}). A1689 is an interesting case, as it is not part of the main subsample since it has a  thermal stability parameter much higher than 5, and it has a radio halo \citep{vacc11}. This system may have been heated up by a major merger \citep{ande04} that did not destroy the cool core. 
 
\subsection{Duty Cycle of Radio Mode Feedback\label{results}}
As discussed in Section~\ref{S:duty_cycle_results}, our simulations imply an AGN heating duty cycle of at least $\sim 60$\% and up to 100\%. This result implies that cooling flow systems do not undergo long periods of time in which no significant cavities are present in their hot atmospheres. Therefore, it appears from our results that the atmospheres of cooling flows are undergoing almost non-stop energy injection through bubbles inflated by the central AGN. The exact nature of this heating remains unclear \citep[for a review, see][]{mcna07}, but our results clearly point to a connection between the presence of gas that is unstable to cooling (traced by low values of $\eta_{\rm min}$) and the bubbles. This connection supports the idea that cold gas accretion, which might be induced by mergers or cooling of the ICM, fuels the outbursts that create the bubbles instead of hot accretion (i.e., Bondi accretion).

The heating duty cycle that we have derived is distinct from the jet duty cycle, which is the fraction of time that the radio jet is active. Our results do not address the jet duty cycle, which, when traced only by the bubble population, could range from small values if the jets operate in short, intermittent bursts (which inflate the cavities and then stop) to values approaching unity if instead the jets are continuous (and the bubbles detach from them as they rise). However, we found that almost every source with cooling time below $\sim 5 \times 10^{8}$ yr has radio lobes. Our results should be useful as constraints on (or inputs to) simulations that include radio-mode AGN heating.

\subsection{Cooling Flow Systems without Bubbles\label{disc_cav}}
 
Figure \ref{F:tcool_instab_vs_lrad} left shows that the systems with bubbles  have short cooling times ( $\lesssim 5 \times 10^{8}$ yr) and the majority of them have high radio luminosities ($\gtrsim 2.5 \times$ 10$^{30}$ erg s$^{-1}$ Hz$^{-1}$). However, as discussed in the previous section, some systems with short cooling times do not appear to have bubbles \citep[e.g.,][]{raff08}. Here we further examine these systems to understand whether they are likely to have bubbles that are simply missed in observations. In particular, we examine their radio properties, star formation properties and merger activity. 

When the radio luminosity is considered, we find that many of the objects with high radio luminosity ($\gtrsim 2.5 \times$ 10$^{30}$ erg s$^{-1}$ Hz$^{-1}$)
that do not show bubbles in the existing \emph{Chandra} images are ``coronae'' \citep{sun07,sun09}. In this category are A400, A3376, A2634, A3395s and Coma. The only coronae from this category for which we found bubbles are A3391 and A3532. However, all these coronae have large radio lobes \citep{owen85,bagc06,hard05,bock99,beck95,vent01}. The interaction between these radio lobes and the ICM is expected to create bubbles, as in A3391 and A3532, 
but deeper X-ray images are needed to detect any bubbles embedded in the low density gas of these systems. Another object with high radio luminosity, short cooling time and no visible bubbles in the existing \emph{Chandra} image is EXO 0423.4-0840, which has a radio source at the centre of the cluster with very small lobes \citep{bels05}. Again, the interaction between these lobes and the ICM might create bubbles, but we are not able to see the bubbles because the \emph{Chandra} point spread function is too large to resolve them.  

The remaining category of objects that have high radio luminosity, but lack bubbles, are massive clusters. This may be the case for  A1644, A3112 and RXCJ 1504.1-0248. Except  possibly A3112 \citep{taki03}, these systems lack radio lobes. Therefore, these systems may well lack substantial bubbles, with the possible exception of A3112, which has the shortest cooling time and some hints of radio lobes. 

These systems may be in a cooling stage, between AGN outbursts. \citet{binn95} postulated that there may be short bursts of nuclear activity alternating with periods of quiescent cooling. While A3112 shows very little star formation \citep{hick10}, RXCJ 1504.1-0248 has a very high star formation rate \citep{ogre10,raff08} and a mini-halo \citep{giac11} most probably related to the cooling flow activity, as in Perseus \citep{gitt02}. A1644 also has high rates of star formation \citep{craw99,odea10}. Furthermore, sloshing  might provide enough heat to balance the inner cooling of the gas \citep{zuho10,zuho11}. This may be the case for A1644 which is thought to be undergoing a merger between the main cluster and a subcluster \citep{reip04,john10}. 

The cooling flow systems with a radio luminosity below 2.5 $\times$ 10$^{30}$ erg s$^{-1}$ Hz$^{-1}$ are mostly ellipticals, groups or poor clusters (e.g., NGC 1550, NGC 4636, NGC 507, NGC 5044, Fornax, MKW 4, AWM 7, A262, A133) and most of them show bubbles in the existing \emph{Chandra} images, except AWM 7, which is probably undergoing a merger \citep{pele90}, and MKW 4. Both AWM 7 and MKW 4 do not have any lobe radio emission, but they have star formation activity \citep{mcna89}, and thus they might be in a cooling stage, as proposed above for the more massive clusters like RXCJ 1504.1-0248. 

Another category of cooling flow objects with low radio luminosity are the corona class systems: A1060, III ZW 54, 3C 129.1, A3558, A2589 and MKW 8. For A1060 and III Zw 54 we measured bubbles (see Section \ref{Cav_sect}), and 3C 129.1 has extended radio emission, and thus might have bubbles that are not visible in the existing shallow \emph{Chandra} image. On the other hand, MKW 8, A2589 and A3558 do not have extended radio lobes (see Table \ref{radio_info_table}). 

There are also massive clusters with low radio luminosity: Ophiuchus, A1650 and A2065. The central galaxies in these clusters only show unresolved nuclear radio emission and no star formation, but A2065 and Ophiuchus might be in the midst of mergers (see Table \ref{radio_info_table}). Also Ophiuchus has a radio mini-halo \citep{govo09,murg10}, and shows evidence of sloshing in its core \citep{mill10}, which might be related \citep{zuho11b}. As in the case of RXCJ 1504.1-0248, the mini-halo in Ophiuchus might be related to the previous cooling flow activity, and the sloshing has revived this old radio emission.  Furthermore, \citet{ecke08} and \citet{neva09} show evidence of possible hard X-ray emission, which might be of non-thermal origin, caused by Compton scattering of cosmic microwave background radiation by the same population of relativistic electrons responsable for the mini-halo emission. As a result, Ophiuchus, A1650 and A2065 might be similar to the intermediate class systems, those with a cooling time between  $\sim 5 -10 \times 10^{8}$ yr, which presently are not expected to have bubbles since they are heated by major mergers, but may previously have had cooling flow activity and bubbles (see Section  \ref{disc_cf_vs_ncf}).

Lastly, we note that small bubbles  may be present in these systems but missed in current observations. Most of the B55 and HIFLUGCS systems are at moderate redshifts (around $z=0.05$), with a few systems above a redshift of 0.1 (A1689 and RXC J1504.1-0248). Therefore, bubbles such as those seen in M87 \citep[e.g.,][]{form05} would be too small to image. For example, if we extrapolate one of the inner cavities from M87, with a radius of 12 arcsec \citep{form05}, to a higher redshift of 0.05 we will obtain a radius of 1 arcsec. As a result, this kind of bubble would be completely missed in current X-ray observations.

\subsection{Cooling Flow Versus Non-Cooling Flow Systems\label{disc_cf_vs_ncf}} 
 
The CF/NCF dichotomy has been studied a great deal recently using observations \citep[e.g.,][]{sander06,sander09,chen07,cava09,prat09,huds10} and simulations \citep[e.g.,][]{pool08,burn08,mcca08,guo09}. Using a statistically selected sample, \citet{sander09} found that the dichotomy seems to be real and not an archival bias. There are two main underlying origins for the CF/NCF dichotomy. The first assumes that the separation occurs early due to pre-heating \citep{mcca08} through mergers \citep{pool08,burn08} or from TeV gamma-rays from blazars \citep{pfro11}. The second interpretation assumes that the separation occurs late, such that a CF cluster can be destroyed due to merger \citep{ross10,ross11} or powerful AGN outburst \citep{guo09}. However, \citet{pfro11} argue that the AGN heating is likely insufficient to turn a CF into a NCF cluster, but the impact of AGN induced turbulence could result into a NCF atmosphere on time scale larger than 1 Gyr \citep{parr10,rusz10}.

From Figures \ref{F:tcool_instab_vs_lrad} (left panel) and \ref{F:sep_tcool}, we found that clusters are separated into two different groups based on the central cooling time, central source radio luminosity and the separation between the X-ray peak and the optical centroid. Clusters with low central cooling times ($\lesssim 5 \times 10^{8}$ yr) and smaller separation ( $\lesssim 12$ kpc) have a cooling flow activity and  many of them show bubbles in their atmospheres. Clusters with longer cooling times ($\gtrsim 10^{9}$ yr), low central radio luminosities ($\lesssim 2.5 \times$ 10$^{30}$ erg s$^{-1}$ Hz$^{-1}$) and larger separations (above $12$ kpc) often show large haloes and relics likely due to merger shocks.
 
For the cooling flow systems, most have visible bubbles, or have extended radio emission from the central source (e.g., A3112), or appear to be in a cooling stage (e.g., A1644 and RXCJ 1504.1-0248). The systems with a cooling time between $\sim 5 \times 10^{8}$~yr and $10^{9}$ yr (A576, A2244, A3571, A2142, A2065, A2063, A1651, A4038) do not have visible bubbles. This result agrees with the findings of \citet{raff08}, since they are above the threshold for the onset of star formation (and hence significant cooling is probably not occurring). For the central radio source, none of them have more than point-like central radio emission (see Table \ref{radio_info_table}), except A2063 which has small radio lobes and possible cavities \citep{kano06}. Due to the fact that most of them are experiencing some sort of merger activity, diffuse emission might be also present in some of them, such as A2142, A3571, and A4038 (see Table \ref{radio_info_table} for references and M. Rossetti private communication for the detection of a radio halo in A2142).

Among the 8 clusters with intermediate cooling times,  only A2063 and A1651 lack the central temperature decrease typical of cooling flow clusters (see Table \ref{tcool_table} and Section \ref{tcool_Tdrop_instab}). Therefore, these systems share properties of both cooling flow and non cooling flow systems, and thus may represent an intermediate class of systems which are presently heated by a major merger but previously might have had a cooling flow activity and bubbles. However, one cannot say whether they will evolve toward non-cooling flow as \citet{ross11} have suggested, or if they might restart the cycle of cooling and heating again at some point. 

Lastly, for the objects with central cooling times above $\sim  10^{9}$ yr, none show any extended emission for the central radio source, except A2147 which shows small lobes (see Table \ref{radio_info_table}), and they do not have visible bubbles A2  147 might have been a corona system which was destroyed by the merger \citep{sun07}. However, this process has low probability even for corona systems \citep{sun07}. All these systems with a central cooling time above $\sim  10^{9}$ yr might be experiencing majors mergers due to the continuing growth of large-scale structure. Therefore, they are classified as non-cooling flow systems for which one does not expect the presence of bubbles. 
 
We can conclude that cooling flow systems (including the intermediate systems) and non-cooling flow systems are well separated, expecially when we look at the central cooling time versus the separation between the X-ray peak and the optical core (see Figure \ref{F:sep_tcool}). Furthermore, the intermediate systems separate from the cooling flow systems, with a central cooling time below $\sim 5 \times 10^{8}$ yr, if we look at the central radio luminosity (see Figure \ref{F:tcool_instab_vs_lrad} left) and the presence of an extended central radio source (see Table \ref{radio_info_table}). The intermediate systems have a radio luminosity below 2.5 $\times$ 10$^{30}$ erg s$^{-1}$ Hz$^{-1}$ and they do not have extended central radio sources. However, from observations there is no conclusive way to know if a cluster like A2256 will become a cooling flow system (or if an object like A2065 will become a non cooling flow one).
 
The merging NCF systems and the CF systems are quite separate in Figure \ref{F:tcool_instab_vs_lrad}. Thus any transitions between the two types must be rare and/or fast. However, a transition from CF to NCF caused by a major merger or a very powerful AGN outburst was found to be improbable \citep{pool08,pfro11}. Transitions in the opposite direction could be the result of gas cooling or the injection of low entropy gas during a merger. However, the speed of the former process is determined by the cooling time of the central gas, while it remains unclear whether the latter process is possible since the turbulence at the cluster centre is sustained for several gigayears after a cluster merger \citep{paul11}. Further simulations are required to establish which, if any, of these transitions are physically plausible and could be consistent with the data.

\section{Conclusions}
The duty cycle of radio mode feedback was determined for two overlapping, complete samples of cooling flow clusters: B55 and HIFLUGCS. All members of both samples have been observed with \emph{Chandra}. As a result, these observations allow us with certainty to put limits on the non-detection of bubbles in some systems and to detect more bubbles in others. We found bubbles in 7 systems that were not previously reported (NGC 507, III Zw 54, NGC 1550, A496, A3391, A1060 and A3532). 

We identified a total of 49 cooling flow systems (32 for B55 and 41 for HIFLUGCS) using two criteria:  $\eta_{\rm min}< 5$ and X-ray to optical core separation $<$ 12 kpc. However, we caution that $\eta_{\rm min}$ is relevant to AGN heating only if the AGN is fueled by cooled gas (as opposed to accretion of hot gas, i.e., Bondi accretion). Our results imply that $\sim 60\%$ of all clusters are cooling flows: $\sim 57$\% for B55 and $\sim 64$\% for HIFLUGCS. These numbers agree well with the result of \citet{huds10} for HIFLUGCS of $72 \%$. If we exclude the corona systems (13 in total, 7 for B55, and 11 for HIFLUGCS), the percentage of cooling flow systems is $\sim 45\%$ ($\sim 45$\% for B55 and $\sim 47$\% for HIFLUGCS). Out of 49 cooling flow systems, 31 have bubbles (22 for B55, and 26 for HIFLUGCS). There are 7 intermediate cooling flow systems (7 for B55, and 6 for HIFLUGCS), and none of them have bubbles. Therefore, the duty cycle of radio-mode feedback is at least $\sim 69\%$ for B55 and $\sim 63\%$ for HIFLUGCS. If we exclude the corona class systems, we found a duty cycle of $\sim 76\%$ for B55 and $\sim 77\%$ for HIFLUGCS.

We found that the more luminous central radio sources (those with $L_{1400} \gtrsim 2.5 \times 10^{30}$ erg s$^{-1}$ Hz$^{-1}$) are only found in cooling flows (as are the X-ray bubbles). This result supports the AGN feedback scenario, in which cooling from the hot atmosphere results in AGN outbursts (traced by their luminous radio emission) that in turn heat the gas and prevent periods of intense, long-duration cooling. 

However, some bubbles could be missed in existing observations. For the systems without detected bubbles, we used simulations to limit the energy that could be present in undetected bubbles, finding that the power being injected by the central AGN into most of these systems could balance cooling, although the bubbles remain undetected. Therefore, with the existing data, one can not exclude the possibility that all cooling flow clusters have bubbles with enough power to balance cooling, implying a duty cycle of up to $\sim 100\%$. This result needs to be investigated further, especially using radio observations, to see if all cooling flow clusters have radio lobes that might create bubbles.

\section*{Acknowledgments}
LB's work at Pennsylvania State University was supported by \emph{Chandra} X-ray grant AR9-0018x. PEJN's work was supported in part by NASA grant NAS8-03060. LB thanks George Pavlov and Niel Brandt for their support at Pennsylvania State University and J. Nevalainen and D. Eckert for useful discussions about the Ophiuchus Cluster.

\bibliographystyle{mn2e}
\bibliography{/Users/birzan/Documents/Bibliography/master_references}

\begin{table*}
\footnotesize
\begin{minipage}{168mm}
  \caption{X-ray and Optical Coordinates.}
\begin{tabular}{@{}lccccccc} 
   &  & \multicolumn{2}{c}{X-Ray Core (J2000)}$^a$ &   &\multicolumn{2}{c}{CDG Core (J2000)}$^c$ &  $\Delta r$$^d$\\
\cline{3-4} \cline{6-7}
 System & z & RA & DEC & CDG Name$^b$ & RA & DEC & (kpc) \\
 \hline
A85               & 0.055 & 00 41 50.2  & -09 18 11  & PGC 002501                                  & 00 41 50.5 & -09 18 11 & 4.8  	\\
A119              & 0.044 & 00 56 15.9* & -01 15 13* & PGC 073498                                  & 00 56 16.1 & -01 15 20 & 6.7  		\\
A133              & 0.057 & 01 02 41.7  & -21 52 48  & ESO 541-013                                 & 01 02 41.7 & -21 52 55 & 8.3  		\\
NGC 507           & 0.017 & 01 23 39.9  & +33 15 21  & NGC 507*                  				   & 01 23 39.9 & +33 15 22 & 0.2  			 			\\
A262              & 0.016 & 01 52 46.3  & +36 09 12  & NGC 708                                     & 01 52 46.5 & +36 09 07 & 1.8  		\\
AWM 7             & 0.017 & 02 54 27.6  & +41 34 43  & NGC 1129                                    & 02 54 27.3 & +41 34 47 & 1.4  		\\
A400              & 0.024 & 02 57 41.7  & +06 01 36  & NGC 1128*                   	  			   & 02 57 41.7 & +06 01 35 & 0.3  			 			\\
A399              & 0.072 & 02 57 53.3* & +13 01 31* & UGC 02438                                   & 02 57 53.2 & +13 01 52 & 29.2 		\\
A401              & 0.074 & 02 58 58.1* & +13 35 00* & UGC 02450                                   & 02 58 57.9 & +13 34 59 & 3.9  		\\
A3112             & 0.075 & 03 17 57.6  & -44 14 17  & ESO 248-G 006                               & 03 17 57.7 & -44 14 18 & 1.4  		\\
Perseus           & 0.018 & 03 19 48.1  & +41 30 44  & NGC 1275                                    & 03 19 48.2 & +41 30 42 & 0.9  		\\
NGC 1399          & 0.005 & 03 38 29.1  & -35 27 01  & NGC 1399*                 				   & 03 38 28.9 & -35 27 01 & 0.2  			 			\\
2A 0335+096       & 0.035 & 03 38 41.1  & +09 57 59  & PGC 013424                                  & 03 38 40.6 & +09 58 12 & 11.2 		\\
III Zw 54         & 0.029 & 03 41 17.6  & +15 23 48  & 2MASX J0341175+152348*      				   & 03 41 17.5 & +15 23 48 & 0.5  			 			\\
A3158             & 0.060 & 03 42 56.8* & -53 37 47* & PGC 013641*                 				   & 03 42 53.0 & -53 37 52 & 39.8 			 			\\
A478              & 0.088 & 04 13 25.3  & +10 27 55  & PGC 014685                                  & 04 13 25.3 & +10 27 55 & 1.0  		\\
NGC 1550          & 0.012 & 04 19 38.0  & +02 24 35  & PCG 014880                                  & 04 19 38.0 & +02 24 35 & 0.1  		\\
EXO 0422-086      & 0.040 & 04 25 51.3  & -08 33 38  & MCG-01-12-005*              				   & 04 25 51.3 & -08 33 39 & 0.6  			 			\\
A3266             & 0.059 & 04 31 14.4* & -61 27 22* & ESO 118-IG 030*             				   & 04 31 13.4 & -61 27 12 & 13.3 			 			\\
A496              & 0.033 & 04 33 37.8  & -13 15 43  & PGC 015524                                  & 04 33 37.8 & -13 15 43 & 0.3  		\\
3C 129.1          & 0.022 & 04 50 06.6  & +45 03 06  & PGC 016124                                  & 04 50 06.7 & +45 03 06 & 0.5  	\\
A3376             & 0.046 & 06 02 09.7* & -39 57 00* & PGC 018297*                 				   & 06 02 09.7 & -39 57 01 & 0.7  				 			\\
A3391             & 0.051 & 06 26 20.5  & -53 41 36  & ESO 161-IG 007*             				   & 06 26 20.5 & -53 41 36 & 0.4  			 			\\
A3395s            & 0.051 & 06 26 49.6  & -54 32 35  & PGC 019057                                  & 06 26 49.6 & -54 32 35 & 0.5  		\\
A576              & 0.039 & 07 21 30.4  & +55 45 42  & CGCG 261-056 NED05*          			   & 07 21 30.3 & +55 45 42 & 1.7  			 			\\
PKS 0745-191      & 0.103 & 07 47 31.4  & -19 17 41  & PGC 021813                                  & 07 47 31.3 & -19 17 40 & 2.1  		\\
A644              & 0.070 & 08 17 25.7* & -07 30 34* & 2MASX J08172559-0730455                     & 08 17 25.6 & -07 30 46 & 16.0 		\\
A754              & 0.054 & 09 09 18.4* & -09 41 19* & 2MASX J09091111923-0941591                  & 09 09 19.1 &- 09 42 03 & 47.3 		\\
Hydra A           & 0.055 & 09 18 05.7  & -12 05 44  & PGC 026269                                  & 09 18 05.7 & -12 05 44 & 0.5  		\\
A1060             & 0.013 & 10 36 42.8  & -27 31 42  & NGC 3311*                   				   & 10 36 42.7 & -27 31 41 & 0.3  			 			\\
A1367             & 0.022 & 11 44 44.0* & +19 42 28* & NGC 3862*                   				   & 11 45 05.0 & +19 36 27 & 208  			 			\\
MKW 4             & 0.020 & 12 04 27.2  & +01 53 46  & NGC 4073                                    & 12 04 27.0 & +01 53 45 & 1.3  		\\
ZwCl 1215         & 0.075 & 12 17 41.3* & +03 39 37* & PGC J1217+0339*             				   & 12 17 41.1 & +03 39 21 & 22.9 			 			\\
M87               & 0.004 & 12 30 49.4  & +12 23 28  & M87                                         & 12 30 49.4 & +12 23 28 & 0.04 		\\
NGC 4636          & 0.003 & 12 42 50.0  & +02 41 12  & NGC 4636                                    & 12 42 49.8 & +02 41 16 & 0.3  		\\
Centaurus         & 0.011 & 12 48 48.9  & -41 18 40  & NGC 4696                                    & 12 48 49.3 & -41 18 40 & 1.0  		\\
A1644             & 0.047 & 12 57 11.8  & -17 24 32  & 2MASX J12571157-1724344                     & 12 57 11.6 & -17 24 35 & 3.3  		\\
A3532             & 0.055 & 12 57 22.1  & -30 21 47  & 2MASX J12572091-3021245*    				   & 12 57 21.9 & -30 21 48 & 2.8       \\  			 			A1650             & 0.084 & 12 58 41.5  & -01 45 42  & PGC 1110773                                 & 12 58 41.5 & -01 45 41 & 2.3  		\\
A1651             & 0.085 & 12 59 22.4* & -04 11 40* & 2MASX J12592251-0411460                     & 12 59 22.5 & -04 11 45 & 8.5  		\\
Coma              & 0.023 & 12 59 35.6  & +27 57 33  & NGC 4874                                    & 12 59 35.7 & +27 57 33 & 0.7  		\\
A1689             & 0.183 & 13 11 29.5  & -01 20 26  & MaxBCG J197.87292-01.34110* 				   & 13 11 29.4 & -01 20 28 & 6.2  			 			\\
NGC 5044          & 0.009 & 13 15 24.0  & -16 23 08  & PGC 046115                                  & 13 15 23.8 & -16 23 08 & 0.5  		\\
A1736             & 0.046 & 13 26 49.8* & -27 10 15* & ESO 509-G 009                               & 13 26 48.6 & -27 08 35 & 90.7 		\\
A3558             & 0.048 & 13 27 56.9  & -31 29 44  & ESO 444-G 046                               & 13 27 56.8 & -31 29 45 & 1.3  		\\
A3562             & 0.049 & 13 33 37.7* & -31 40 14* & ESO 444-G 072                               & 13 33 34.7 & -31 40 21 & 36.9 		\\
A3571             & 0.039 & 13 47 28.1  & -32 52 01  & ESO 383-G 076                               & 13 47 28.4 & -32 51 54 & 6.5  		\\
A1795             & 0.063 & 13 48 52.7  & +26 35 29  & PGC 049005                                  & 13 48 52.5 & +26 35 34 & 7.9  		\\
PKS 1404-267      & 0.023 & 14 07 29.8  & -27 01 05  & IC 4374                                     & 14 07 29.7 & -27 01 04 & 0.5  		\\
MKW 8             & 0.027 & 14 40 42.8  & +03 27 57  & NGC 5718*                   				   & 14 40 42.8 & +03 27 56 & 0.6  			 			\\
RXCJ 1504.1-0248  & 0.215 & 15 04 07.5  & -02 48 17  & LCRS B150131.5-023636                       & 15 04 07.5 & -02 48 17 &  3.5 		\\
A2029             & 0.077 & 15 10 56.1  & +05 44 41  & PGC 054167                                  & 15 10 56.1 & +05 44 41 & 0.8  	\\
A2052             & 0.035 & 15 16 44.4  & +07 01 17  & UGC 09799                                   & 15 16 44.5 & +07 01 18 & 0.9  		\\
MKW 3s            & 0.045 & 15 21 51.7  & +07 42 23  & NGC 5920                                    & 15 21 51.8 & +07 42 32 & 7.8  		\\
A2065             & 0.073 & 15 22 29.2  & +27 42 26  & PGC 054888                                  & 15 22 28.9 & +27 42 45 & 3.3  		\\
A2063             & 0.035 & 15 23 05.3  & +08 36 37  & CGCG 077-097                                & 15 23 05.3 & +08 36 33 & 3.2  	\\
A2142             & 0.091 & 15 58 20.2  & +27 13 57  & 2MASX J15582002+2714000                     & 15 58 20.0 & +27 14 00 & 7.3  		\\
A2147             & 0.035 & 16 02 15.4* & +15 58 10* & UGC 10143                                   & 16 02 17.0 & +15 58 29 & 20.6 	\\
A2163             & 0.203 & 16 15 46.8* & -06 08 39* & 2MASX J16154140-0609076*    				   & 16 15 41.4 & -06 09 07 & 283  			 			\\
A2199             & 0.030 & 16 28 38.0  & +39 33 04  & NGC 6166                                    & 16 28 38.5 & +39 33 06 & 3.6  		\\
\hline
\end{tabular}
\end{minipage}
\end{table*}

\begin{table*}
\begin{minipage}{126mm}
\contcaption{}
\label{info_table}
\begin{tabular}{@{}lccccccc} 
   &  & \multicolumn{2}{c}{X-Ray Core (J2000)}$^a$ &   &\multicolumn{2}{c}{CDG Core (J2000)}$^c$ &  $\Delta r$$^d$\\
\cline{3-4} \cline{6-7}
 System & z & RA & DEC & CDG Name$^b$ & RA & DEC & (kpc) \\
 \hline
A2204             & 0.152 & 16 32 47.1  & +05 34 34  & VLSS J1632.7+0534                           & 16 32 47.0 & +05 34 33 & 2.9  		\\
Triangulum  Aus.      & 0.051 & 16 38 23.4* & -64 21 11* & ESO 101-G 004                               & 16 38 18.1 & -64 21 37 & 43.2 		\\
A2244             & 0.097 & 17 02 42.6  & +34 03 38  & 2MASX J17024247+3403363                     & 17 02 42.5 & +34 03 38 & 2.2  		\\
A2256             & 0.058 & 17 03 59.8* & +78 39 00* & NGC 6331*                   				   & 17 04 26.8 & +78 38 26 & 117  			 			\\
Ophiuchus         & 0.028 & 17 12 27.7  & -23 22 07  & 2MASX J17122774-2322108                     & 17 12 27.7 & -23 22 11 & 2.2  		\\
A2255             & 0.081 & 17 12 41.3* & +64 04 21* & PGC 059830*                				   & 17 12 34.9 & +64 04 15 & 64.5 			 			\\
A2319             & 0.056 & 19 21 13.1* & +43 56 11* & CGCG 230-007 NED03                          & 19 21 10.0 & +43 56 44 & 50.5 		\\
Cygnus A          & 0.056 & 19 59 28.3  & +40 44 02  & Cygnus A                                    & 19 59 28.3 & +40 44 02 & 0.5  		\\
A3667             & 0.055 & 20 12 31.6* & -56 50 24* & IC 4965                                     & 20 12 27.3 & -56 49 36 & 64.6 		\\
Sersic 159/03     & 0.058 & 23 13 58.7  & -42 43 31  & ESO 291-G 009                               & 23 13 58.6 & -42 43 39 & 8.7  		\\
A2589             & 0.041 & 23 23 57.5  & +16 46 37  & NGC 7647                                    & 23 23 57.4 & +16 46 38 & 1.2  		\\
A2597             & 0.085 & 23 25 19.7  & -12 07 27  & PGC 071390                                  & 23 25 19.8 & -12 07 26 & 0.1  		\\
A2634             & 0.031 & 23 38 29.4  & +27 01 53  & NGC 7720*                  				   & 23 38 29.4 & +27 01 53 & 0.1  			 			\\
A2657             & 0.040 & 23 44 55.9* & +09 11 15* & PGC 072804*                 				   & 23 44 57.4 & +09 11 34 & 23.8 	 					\\
A4038             & 0.030 & 23 47 43.3  & -28 08 29  & IC 5358*                   				   & 23 47 45.0 & -45 08 27 & 13.5 			 			\\
A4059             & 0.048 & 23 57 00.4  & -34 45 44  & ESO 349-G 010                               & 23 57 00.7 & -34 45 33 & 10.5 		\\
\hline
\end{tabular}
 \label{info_table}
 \medskip
$^a$The X-ray core positions from this work; the asterisks mark the uncertain core positions (due to the lack of the cooling core or to the presence of a confusing structure).\\
$^b$Notes on some systems (noted with an asterisk). NGC 507 --- this system has a smaller satellite nearby. A400 --- there are two cD galaxies in this system: A400-42, to the north, and A400-43 to the south. There is super-massive binary black hole at the centre of A400 \citep{huds06}, we located the X-ray centre and the CDG centre on the northern AGN. NGC 1399 --- this system has a satellite galaxy. Zw III 54 --- a posible satellite galaxy is nearby. A3158 --- this system has another cD galaxy in the centre part of the cluster, PGC 013652, and the location of the optical centre (PGC 013641) lies in the CCD gap. EXO 0422-086 --- this system has a couple of satellite galaxies. A3266 and A3391 --- the central cD galaxies in these systems appear to be interacting and have an extended corona. A3376 --- this is not the central cD galaxy of the cluster, but the one that corresponds to the centroid of the X-ray emission. A576 --- there is a another, similarly sized cD galaxy in the central part of the cluster, PGC020784. A1060 --- another similarly sized cD galaxy is nearby, NGC 3309. A1367 --- there is another D galaxy, NGC 3842 located north of the X-ray center, which has large scale radio lobes \citep{beck95,owen97}, but neither NGC 3842 or NGC 3862 correspond to the X-ray peak. NGC 3862 hosts also a radio galaxy, 3C 264 \citep{lara99,cond98}. A3532 --- this system has a satellite nearby. A1689 --- there is another equally bright galaxy nearby; the optical image is of poor resolution. MKW 8 --- there is another equally bright galaxy nearby. A2163 --- the identification of the BCG is uncertain because the optical image is of poor resolution. A2256 --- the BCG identification is uncertain since there is another similarly sized cD galaxy in the field (UGC 10726), plus several smaller ones, such as CGCG 355-026 and MCG+13-12-014. Between the four cDs, only MCG+13-12-014 have an associated radio source (in NVSS). A2255 --- there is another similarly sized cD galaxy nearby, PGC 059831. A2634 --- this is a galaxy pair system; the pair is NGC 7720 NED01; A4038 --- there is another smaller cD nearby, PGC 072437.\\
$^c$The CDG core positions are taken from the literature.\\
$^d$The separation between the optical centre and the X-ray center.\\
\end{minipage}
\end{table*}

\begin{table*}
\small
\begin{minipage}{168mm}
  \caption{X-ray Properties.}
\begin{tabular}{@{}lcccccccccc}
System &  Obs.\ ID &  $t$$^a$ & $\Delta T$$^b$ & $t_{\rm cool}$$^c$ &  r$^d$ &  $t_{\rm cool}$ (1 kpc)$^e$ & $\eta_{\rm min}$$^f$ &   $L_{\rm{x}}$$^g$&  $r_{\rm{cool}}$$^h$ \\  
 &   &  (ks) &   &  (Gyr) &  (kpc) &  (Gyr) &   &  (10$^{42}$ erg s$^{-1}$) &  (kpc) \\
\hline
A85              & 904   & 37.9  & $2.30^{+0.31}_{-0.35}$ & $0.20^{+0.09}_{-0.07}$    & 2.6  & $0.15^{+0.07}_{-0.05}$      & 2.38    & $271^{+3}_{-2}$     & 116   \\
A119             & 7918  & 44.0  & \ldots                & $12.7^{+4.1}_{-3.1}$      & 19.7 & \ldots                     & \ldots & \ldots                   & \ldots  \\
A133             & 2203  & 31.4  & $2.54^{+0.56}_{-0.41}$ & $0.28^{+0.04}_{-0.04}$    & 3.5  & $0.12^{+0.04}_{-0.04}$*     & 1.61    & $76^{+1}_{-1}$      & 76     \\
NGC 507          & 2882  & 41.1  & $1.42^{+0.05}_{-0.06}$ & $0.55^{+0.04}_{-0.05}$    & 4.0  & $0.13^{+0.01}_{-0.01}$      & 0.20    & $2.20^{+0.03}_{-0.03}$ & 47     \\
A262             & 7921  & 110.7 & $3.21^{+0.13}_{-0.13}$ & $0.15^{+0.01}_{-0.01}$    & 2.2  & $0.14^{+0.02}_{-0.02}$      & 0.64    & $8.52^{+0.05}_{-0.06}$ & 57     \\
AWM 7            & 908   & 47.9  & $3.34^{+0.15}_{-0.28}$ & $0.09^{+0.03}_{-0.02}$    & 0.8  & $0.13^{+0.03}_{-0.02}$      & 2.63    & $17.6^{+0.2}_{-0.2}$   & 57    \\
A400             & 4181  & 20.5  & \ldots                & $10.5^{+5.4}_{-6.5}$      & 18.8 & $0.6^{+0.3}_{-0.4}$         & 0.83    & $0.47^{+0.03}_{-0.01}$ & 22     \\
A399             & 3230  & 46.5  & \ldots                & $10.5^{+2.6}_{-2.3}$      & 20.9 & $8.4^{+4.0}_{-3.8}$         & \ldots & \ldots                   & \ldots  \\
A401             & 2309  & 11.6  & \ldots                & $7.8^{+3.1}_{-2.9}$       & 24.8 & \ldots                     & \ldots & \ldots                   & \ldots  \\
A3112            & 2516  & 12.3  & $2.47^{+0.15}_{-0.15}$ & $0.14^{+0.02}_{-0.02}$    & 4.4  & $0.06^{+0.01}_{-0.01}$      & 0.09    & $376^{+5}_{-5}$     & 130    \\
Perseus          & 1513  & 17.7  & $2.34^{+0.11}_{-0.11}$ & $0.47^{+0.02}_{-0.02}$    & 8.0  & $ 0.24^{+0.03}_{-0.03}$*    & 0.63    &  $545^{+2}_{-2}$    &   110       \\
NGC 1399         & 9530  & 58.0  & $1.81^{+0.04}_{-0.03}$ & $0.013^{+0.006}_{-0.005}$ & 0.1  & $0.078^{+0.001 }_{-0.001 }$ & 1.04    & $0.39^{+0.01}_{-0.01}$ & 24    \\
2A 0335+096      & 7939  & 49.5  & $3.97^{+0.22}_{-0.19}$ & $0.142^{+0.014}_{-0.013}$ & 2.6  & $0.12^{+0.02}_{-0.02}$      & 0.23    & $166^{+1}_{-1}$     & 57     \\
III Zw 54        & 4182  & 21.2  & $1.38^{+0.20}_{-0.24}$ & $3.5^{+0.7}_{-0.5}$       & 11.7 & $0.12^{+0.02}_{-0.02}$      & 0.05    & $2.48^{+0.07}_{-0.06}$ & 30     \\
A3158            & 3712  & 27.4  & \ldots                & $5.9^{+2.3}_{-1.9}$       & 16.7 & $4.6^{+2.8}_{-2.7}$         & \ldots & \ldots                   & \ldots  \\
A478             & 1669  & 42.4  & $3.16^{+0.21}_{-0.26}$ & $0.19^{+0.01}_{-0.01}$    & 4.5  & $0.06^{+0.02}_{-0.01}$*     & 1.06    & $1209^{+8}_{-9}$    & 161    \\
NGC 1550         & 5800  & 44.5  & $1.37^{+0.08}_{-0.03}$ & $0.065^{+0.004}_{-0.004}$ & 0.8  & $0.082^{+0.004}_{-0.004} $  & 0.91    & $3.38^{+0.04}_{-0.03}$ & 76     \\
EXO 0422-086     & 4183  & 8.9   & $2.54^{+0.78}_{-0.79}$ & $0.36^{+0.18}_{-0.14}$    & 3.3  & $0.27^{+0.02}_{-0.02}$      & 1.98    & $38^{+2}_{-1}$      & 64    \\
A3266            & 899   & 29.8  & $2.04^{+0.73}_{-0.65}$ & $4.6^{+2.1}_{-3.2}$       & 8.4  & $1.6^{+0.8}_{-1.2}$         & 90.0    & \ldots                   & \ldots  \\
A496             & 4976  & 57.5  & $5.65^{+0.30}_{-0.29}$ & $0.10^{+0.01}_{-0.01}$    & 1.3  & $0.09^{+0.02}_{-0.01}$      & 0.89    & $111^{+1}_{1}$   & 87    \\
3C 129.1         & 2219  & 9.6   & $5.0^{+3.0}_{-6.0}$    & $1.8^{+2.3}_{-1.8}$       & 6.6  & $0.3^{+0.3}_{-0.4}$         & 0.75    & $4.9^{0.1}_{-0.2}$     & 24     \\
A3376            & 3202  & 39.3  & $3.83^{+0.87}_{-1.7}$  & $5.7^{+2.8}_{-1.4}$       & 13.9 & $0.4^{+0.2}_{-0.1}$         & 0.25    & $1.0^{+0.1}_{-0.1}$    & 19     \\
A3391            & 4943  & 16.0  & $2.8^{+5.9}_{-3.1}$    & $6.1^{+2.0}_{-2.5}$       & 15.3 & $0.9^{+0.3}_{-0.4}$         & 2.53    & $4.7^{+0.3}_{-0.2}$    & 29    \\
A3395s           & 4944  & 19.8  & $2.4^{+1.0}_{-2.4}$    & $4.5^{+2.7}_{-2.3}$       & 11.9 & $0.10^{+0.06}_{-0.05}$      & 0.04    & $0.8^{+0.1}_{-0.1}$    & 22     \\
A576             & 3289  & 25.6  & $2.3^{+1.0}_{-1.5}$    & $2.6^{+0.8}_{-1.0}$       & 7.4  & $0.7^{+0.2}_{-0.3}$         & 8.7     & $8.1^{+0.2}_{-0.2}$    & 41     \\
PKS 0745-191     & 2427  & 17.9  & $3.02^{+0.48}_{-0.49}$ & $0.24^{+0.08}_{-0.08}$    & 4.7  & $0.19^{+0.08}_{-0.08}$      & 0.68    & $1502^{+15}_{-15}$        & 152      \\ 
A644             & 2211  & 28.9  & \ldots                & $2.4^{+3.2}_{-2.0}$       & 5.6  & \ldots                     & \ldots & \ldots                   & \ldots  \\
A754             & 577   & 43.2  & $1.46^{+1.4}_{-0.57}$  & $11.5^{+2.9}_{-3.1}$      & 20.0 & $8.0^{+5.0}_{-5.0}$         & \ldots & \ldots                   & \ldots  \\
Hydra A          & 4969  & 59.3  & $1.42^{+0.08}_{-0.10}$ & $0.54^{+0.05}_{-0.04}$    & 7.6  & $0.3^{+0.1}_{-0.3}$         & 1.4     & $227{+2}_{-2}$     & 107   \\
A1060            & 2220  & 31.9  & $3.64^{+0.64}_{-0.50}$ & $0.9^{+0.1}_{-0.2}$       & 3.6  & $0.33^{+0.04}_{-0.06}$      & 0.84    &$0.120^{+0.003}_{-0.003}$& 48    \\
A1367            & 514   & 34.4  & \ldots                & $11.7^{+4.2}_{-3.6}$      & 14.7 & \ldots                     & \ldots & \ldots                   & \ldots  \\
MKW 4            & 3234  & 30.0  & $1.75^{+0.21}_{-0.07}$ & $0.12^{+0.01}_{-0.07}$    & 1.5  & $0.09^{+0.01}_{-0.01}$      & 1.80    &  $5.4^{+0.1}_{-0.1}$   & 47     \\
ZwCl 1215        & 4184  & 11.8  & \ldots                & $15.8^{+6.4}_{-5.9}$      & 37.0 & \ldots                     & \ldots & \ldots                   & \ldots  \\
M87              & 3717  & 9.3   & $2.47^{+0.06}_{-0.06}$ & $0.054^{+0.008}_{-0.007}$ & 0.7  & $ 0.08^{+0.08}_{-0.01}$     & 1.22    & $6.3^{+0.1}_{-0.1}$    & 39     \\
NGC 4636         & 323   & 37.0  & $1.57^{+0.05}_{-0.05}$ & $0.027^{+0.003}_{-0.002}$ & 0.2  & $ 0.076^{+0.002}_{-0.002}$  & 0.46    & $0.17^{+0.002}_{-0.002}$ & 61     \\
Centaurus        & 5310  & 49.3  & $5.29^{+0.09}_{-0.09}$ & $0.030^{+0.004}_{-0.003}$ & 0.7  & $ 0.046^{+0.004}_{-0.003}$  &  0.27   & $22.0^{+0.1}_{-0.1}$      & 87    \\
A1644            & 7922  & 51.2  & $2.96^{+0.84}_{-0.61}$ & $0.70^{+0.14}_{-0.13}$    & 6.6  & $0.33^{+0.08}_{-0.08}$      & 1.08    & $17.3^{+0.3}_{-0.3}$   & 55     \\
A3532            & 10745 & 9.4   & $2.10^{+0.82}_{-0.82}$ & $8.1^{+1.4}_{-1.7}$       & 23.3 & $0.24^{+0.04}_{-0.05}$      & 0.13    & $2.2^{+0.3}_{-0.3}$    & 22     \\
A1650            & 4178  & 25.6  & $1.62^{+0.27}_{-0.28}$ & $1.1^{+0.4}_{-0.3}$       & 8.1  & $0.28^{+0.11}_{-0.09}$      & 3.17    & $234^{+3}_{-3}$     & 97     \\
A1651            & 4185  & 8.9   & \ldots                & $4.0^{+3.0}_{-3.0}$       & 6.2  & $1.0^{+0.8}_{-0.9}$         & 21.4    & $150^{+2}_{-3}$     &  94    \\
Coma             & 1086  & 9.5   & $10.0^{+20}_{-5.0}$  & $0.7^{+0.3}_{-0.3}$         & 3.1  & $0.12^{+0.05}_{-0.06}$      & 0.33    & $4.0^{+0.1}_{-0.1}$    & 20     	\\
A1689            & 7289  & 64.4  & $1.81^{+0.43}_{-0.48}$ & $1.0^{+0.3}_{-0.2}$       & 7.6  & $0.6^{+0.2}_{-0.1}$         & 18.9    & $1333^{+15}_{-13}$        & 119    \\
NGC 5044         & 9399  & 81.9  & $2.04^{+0.08}_{-0.08}$ & $0.049^{+0.003}_{-0.002}$ & 0.5  & $ 0.088^{+0.003}_{-0.003}$  & 0.22    & $4.0^{+0.02}_{-0.02}$     & 50     \\
A1736            & 4186  & 11.4  & \ldots                & $14.0^{+32.0}_{-12.0}$    & 40.0 & \ldots                     & \ldots & \ldots                   & \ldots  \\
A3558            & 1646  & 11.2  & $2.57^{+0.76}_{-0.71}$ & $2.0^{+0.5}_{-0.5}$       & 9.0  & $0.17^{+0.08}_{-0.05}$      & 0.39    & $75^{+2}_{-1}$      & 75     \\
A3562            & 4167  & 18.0  & $1.24^{+0.36}_{-0.44}$ & $3.4^{+0.8}_{-0.8}$       & 9.7  & \ldots                     & \ldots & \ldots                   & \ldots  \\
A3571            & 4203  & 31.0  & $2.0^{+0.31}_{-0.37}$  & $1.1^{+0.4}_{-0.3}$       & 4.2  & $0.6^{+0.7}_{-0.7}$         & 59.7    & $88^{+1}_{-1}$   & 55     \\ 
A1795            & 3666  & 14.4  & $2.62^{+0.39}_{-0.38}$ & $0.4^{+0.1}_{-0.1}$       & 1.2  & $0.4^{+0.2}_{-0.2}$         & 3.3     & $510^{+5}_{4}$      & 116    \\
PKS 1404-267     & 1650  & 7.1   & $1.98^{+0.29}_{-0.28}$ & $0.44^{+0.09}_{-0.08}$    & 5.0  & $0.3^{+0.1}_{-0.1}$         & 0.73    & $18^{+1}_{-1}$      & 69    \\
MKW 8            & 4942  & 22.9  & \ldots                & $12.5^{+3.4}_{-3.0}$      & 23.0 & $0.4^{+0.1}_{-0.1}$         & 0.40    & $0.27^{+0.02}_{-0.01}$ & 19     \\
RXCJ 1504.1-0248 & 5793  & 36.6  & $2.71^{+0.32}_{-0.30}$ & $0.23^{+0.13}_{-0.08}$    & 3.4  & $0.2^{+0.1}_{-0.1}$         & 0.47    & $4080^{+49}_{-55}$        &  172   \\
A2029            & 4977  & 77.8  & $3.28^{+0.22}_{-0.25}$ & $0.06^{+0.01}_{-0.01}$    & 1.0  & $0.06^{+0.01}_{-0.01}$      & 3.84    & $1049^{+5}_{-5}$      & 127   \\
A2052            & 890   & 36.8  & $5.06^{+0.58}_{-0.31}$ & $0.6^{+0.5}_{-0.2}$       & 2.8  & $0.3^{+0.2}_{-0.1}$         &  0.42   &  $97^{+1}_{-1}$     & 80     \\
MKW 3s           & 900   & 55.3  & $3.23^{+0.35}_{-0.40}$ & $0.6^{+0.4}_{-0.2}$       & 3.3  & $0.3^{+0.2}_{-0.1}$         & 4.06    & $104^{+1}_{-1}$     & 92    \\
A2065            & 3182  & 21.8  & $1.89^{+0.54}_{-0.44}$ & $2.2^{+0.50}_{-0.5}$      & 14.6 & $0.7^{+0.2}_{0.2}$          & 2.68    & $63^{+1}_{1}$       & 63     \\
A2063            & 6263  & 16.8  & \ldots                & $2.3^{+1.0}_{-1.4}$       & 7.2  & $1.0^{+0.6}_{-0.5}$         & 18.9    & $37^{+1}_{1}$    & 68    \\
A2142            & 5005  & 44.6  & $1.97^{+0.72}_{-1.0}$  & $1.8^{+0.5}_{-0.6}$       & 6.7  & $0.8^{+0.4}_{-0.4}$         & 16.2    & $749^{+5}_{-5}$     & 126    \\
A2147            & 3211  & 17.4  & $2.44^{+1.0}_{-0.62}$  & $10.0^{+2.5}_{-3.5}$      & 20.0 & $1.9^{+0.6}_{-0.8}$         & 6.0     & \ldots                   & \ldots  \\
\hline
\end{tabular}
\end{minipage}
\end{table*}

\begin{table*}
\begin{minipage}{168mm}
\contcaption{}
\label{tcool_table}
\begin{tabular}{@{}lcccccccccc}
System &  Obs.\ ID &  $t$$^a$ & $\Delta T$$^b$ & $t_{\rm cool}$$^c$ &  r$^d$ &  $t_{\rm cool}$ (1 kpc)$^e$ & $\eta_{\rm min}$$^f$ &   $L_{\rm{x}}$$^g$&  $r_{\rm{cool}}$$^h$ \\  
 &   &  (ks) &   &  (Gyr) &  (kpc) &  (Gyr) &   &  (10$^{42}$ erg s$^{-1}$) &  (kpc) \\
\hline
A2163            & 1653  & 67.4  & \ldots                & $12.4^{+5.4}_{-5.5}$      & 45.0 & $6.6^{+3.1}_{-3.1}$         & \ldots & \ldots                   & \ldots  \\
A2199            & 498   & 15.9  & $2.46^{+0.29}_{-0.27}$ & $0.22^{+0.05}_{-0.05}$    & 2.2  & $0.117^{+0.006}_{-0.005}$*  & 4.79    & $171^{+2}_{2}$      & 106    \\
A2204            & 499   & 10.1  & $3.51^{+0.50}_{-0.55}$ & $0.15^{+0.03}_{-0.03}$    & 4.6  & $0.09^{+0.02}_{-0.02}$      & 0.44    &  $2024^{+31}_{29}$  & 148    \\
Triangulum  Aus. & 1227  & 12.0  & $2.0^{+1.2}_{-1.8}$    & $6.9^{+3.4}_{-3.2}$       & 12.5 & $2.2^{+1.2}_{-1.1}$         & 56.0    & \ldots                   & \ldots  \\
A2244            & 4179  & 56.7  & $1.23^{+0.17}_{-0.21}$ & $1.0^{+0.3}_{-0.3}$       & 5.7  & $0.5^{+0.2}_{-0.1}$         & 12.1    & $316^{+3}_{-3}$     & 113    \\
A2256            & 2419  & 9.2   & \ldots                & $18.7^{+24.7}_{-10.8}$    & 27.4 & \ldots                     & \ldots & \ldots                   & \ldots   \\
Ophiuchus        & 3200  & 50.5  & $5.42^{+0.53}_{-0.37}$ & $0.26^{+0.08}_{-0.09}$    & 3.6  & $0.097^{+0.004}_{-0.005}$   & 0.47    & $324^{+1}_{-2}$     & 83     \\
A2255            & 894   & 36.9  & \ldots                & $15.0^{+1.6}_{-1.9}$      & 33.0 & \ldots                     & 15.2    & \ldots                   & \ldots  \\
A2319            & 3231  & 14.2  & \ldots                & $9.6^{+2.6}_{-1.8}$       & 25.5 & \ldots                     &\ldots  & \ldots                   & \ldots  \\
Cygnus A         & 360   & 33.7  & $2.27^{+0.32}_{-0.33}$ & $0.46^{+0.07}_{-0.06}$    & 7.8  & $ 0.16^{+0.08}_{-0.07}$*    &  2.1    & $293^{+3}_{3}$      & 92     \\
A3667            & 889   & 50.3  & \ldots                & $16.3^{+11.2}_{-7.6}$     & 24.0 & \ldots                     & \ldots & \ldots                   & \ldots  \\
Sersic 159/03    & 11758 & 91.8  & $2.30^{+0.58}_{-0.15}$ & $0.22^{+0.03}_{-0.03}$    & 2.5  & $0.10^{+0.03}_{-0.03}$      & 0.64    & $157^{+1}_{1}$      & 111    \\
A2589            & 7190  & 53.4  & $1.63^{+0.18}_{-0.17}$ & $1.0^{+0.1}_{-0.1}$       & 4.8  & $0.27^{+0.04}_{-0.04}$      & 3.4     & $29^{+1}_{-1}$      & 62    \\
A2597            & 7329  & 53.3  & $2.8^{+1.9}_{-1.1}$    & $0.21^{+0.04}_{-0.03}$    & 3.2  & $ 0.10^{+0.05}_{-0.04}$*    & 0.64    & $367^{+3}_{-3}$     & 119   \\
A2634            & 4816  & 48.5  & $5.0^{+1.6}_{-1.0}$    & $0.77^{+0.04}_{-0.04}$    & 8.36 & $0.13^{+0.09}_{-0.08}$      &  0.12   & $0.6^{+0.4}_{-0.4}$      & 17     \\
A2657            & 4941  & 15.7  & \ldots                & $5.0^{+0.7}_{-1.4}$       & 12.7 & \ldots                   & \ldots & \ldots                   & \ldots  \\
A4038            & 4992  & 29.9  & $1.21^{+0.08}_{-0.07}$ & $1.18^{+0.13}_{-0.13}$    & 5.3  & $1.0^{+1.0}_{-1.0}$         & 9.7     & \ldots                   & \ldots  \\
A4059            & 5785  & 92.1  & $2.20^{+0.08}_{-0.08}$ & $0.57^{+0.11}_{-0.10}$    & 5.3  & $ 0.3^{+0.1}_{-0.1}$*       & 5.4     & $85^{+1}_{-1}$      & 84     \\  
\hline
\end{tabular}
 \label{tcool_table}
 \medskip
$^a$The time on source after reprocessing the data.\\
$^b$The temperature drop, calculated as the ratio between the highest and the lowest temperature of the profile when the profile is rising upward smoothly. In some cases we neglected points that were outliers (e.g., in A576, A3376, A1689). For the systems which do not have a statistically significant temperature drop (e.g., A644) or their temperature profile is rising inward, there is no entry.\\
$^c$The cooling time of the innermost region (in Gyr).\\
$^d$The radius, in kpc, for the innermost region used for spectral deprojection.\\
$^e$The cooling time at 1 kpc derived using the deprojected surface brightness profiles (in Gyr). In some cases the surface brightness deprojection did not work well (due, e.g., to the surface brightness profile dropping or flattening towards the center). In these cases, the cooling time at 1 kpc was extrapolated using the cooling time profile from the single temperature model deprojection. These systems are marked with an asterisk.  For some systems a large extrapolation would be required. In these cases there is no entry for the cooling time at 1 kpc.\\
$^f$Thermal stability parameter from \citet{voit08}; $\rm Inst.=\rm min(kT/ \Lambda n_{e} n_{H} r^{2}$). For some of the systems the instability profile is still rising,  therefore there is no minimum (e.g., A399, A3158, A754, A2163). See text for details.\\
$^g$The X-ray luminosity within the cooling radius (determined from the sum of the deprojected fluxes for regions inside the cooling radius) for the subsample of systems that require heating in order to suppress the inner cooling (see Section \ref{Subsample}).\\
$^h$Cooling radius, defined as the radius within which the gas has a cooling time less than $7.7 \times 10^{9}$ yr.\\
\end{minipage}
\end{table*}

\begin{table*}
\small
\begin{minipage}{168mm}
  \caption{Central Radio Sources, Bubbles, and Mergers}
\begin{tabular}{@{}lccccccccccc} 
System$^a$& Radio$^b$ & $L_{\rm 1.4GHz}$$^c$ & Bubbles & Merger &  Sample$^d$ \\
\hline
A85             & yes(15,91,99)*$^\dagger$$^\dagger$     & 3.98 $\pm$ 0.17 $\times$ 10$^{-2}$ 	   & yes(109)    & yes(37,48,72,129)         & B55, HIFLUGCS		 		 \\
A119            & no(2,40,98)                 & 8.9 $\times$ 10$^{-4}$  			               & no         & yes(62)               & B55, HIFLUGCS 					 \\
A133            & yes(11,110,126)*$^\dagger$$^\dagger$ & 1.8 $\pm$ 0.1 $\times$ 10$^{-2}$ 	   &  yes(109)   & yes(110)               & HIFLUGCS 						 \\
NGC 507         & yes(30,51,95,105)*         & 7.15 $\pm$ 0.35 $\times$ 10$^{-3}$ 	           & yes(31,143) & yes(74,104)            & HIFLUGCS						 \\
A262            & yes(11,22)*               & 4.54 $\pm$ 0.01 $\times$ 10$^{-3}$ 	           & yes(109) & yes(48)               & B55, HIFLUGCS 					 \\
AWM 7           & no(2)                        & 3.3 $\times$ 10$^{-5}$ 				           & no         &  probable(107)         & B55 					 \\
A400*            & yes(65,98)*               & 3.41 $\pm$ 0.21 $\times$ 10$^{-1}$ 	           & no         & yes(65)               & HIFLUGCS 						 \\
A399            & no(2,94)$^\dagger$              & 1.84 $\times$ 10$^{-3}$ 				   & no         & yes(94,116)               & B55, HIFLUGCS 					 \\
A401            & no(2)$^\dagger$              & 1.94 $\times$ 10$^{-3}$ 				       & no         & yes(94,116)               & B55, HIFLUGCS 					 \\
A3112           & yes(91,128)*              & 1.72 $\pm$ 0.012 					               & no         & \ldots               & B55, HIFLUGCS 					 \\
Perseus         & yes(11,38)*               & 1.70 $\pm$ 0.07 						           & yes(109) & yes(24,48)               & B55 						 \\
NGC 1399        & yes(2,124)*                 & 2.91 $\pm$ 0.01 $\times$ 10$^{-3}$ 	           &  yes(124)   & \ldots               & HIFLUGCS 						 \\
2A 0335+096     & yes(11,118,122)*           & 1.06 $\pm$ 0.05 $\times$ 10$^{-2}$ 	           & yes(109)    & yes(48,142)            & B55, HIFLUGCS 				 \\
III Zw 54*       & yes(2,91)                   & 3.85 $\pm$ 0.19 $\times$ 10$^{-3}$ 	           & yes(143)    & \ldots               & HIFLUGCS 						 \\
A3158           & yes(68)                   & 9.11 $\pm$ 0.12 $\times$ 10$^{-3}$ 	           & no         & yes(141)               & B55, HIFLUGCS 					 \\
A478            & yes(11)*                  & 6.95 $\pm$ 0.27 $\times$ 10$^{-2}$ 	           & yes(109)    & \ldots               & B55, HIFLUGCS 					 \\
NGC 1550        & yes(34)*                  & 5.47 $\pm$ 0.53 $\times$ 10$^{-4}$ 	           & yes(31,143) & yes(71)               & HIFLUGCS 						 \\
EXO 0422-086    & yes(9)*                   &  4.1 $\pm$ 0.1 $\times$ 10$^{-2}$ 	           & no         & \ldots               & HIFLUGCS 						 \\
A3266           & no(2)                        & 1.46 $\times$ 10$^{-3}$ 				       & no         & yes(43,48,63,121)      & B55, HIFLUGCS 			 \\
A496            & yes(2)                    & 2.96 $\pm$ 0.11 $\times$ 10$^{-2}$ 	           & yes(143)    & yes(35,48,114,133)         & B55, HIFLUGCS 			 \\
3C 129.1*        & yes(75,78,79)*            & 1.00 $\pm$ 0.45 $\times$ 10$^{-4}$ 	           & no         & \ldots               & B55 						 \\
A3376*           & yes(6,91)*$^\dagger$      & 1.91 $\pm$ 0.06 $\times$ 10$^{-1}$ 	           & no         & yes(6,106)             & HIFLUGCS 						 \\
A3391*           & yes(3,56,92,97)*          & 3.99 $\pm$ 0.49 						           &  yes(143)   & yes(33,132)            & B55, HIFLUGCS 				 \\
A3395s*          & yes(3)*                   & 2.05 $\pm$ 0.09 						           & no         & yes(33,132)            & B55, HIFLUGCS 				 \\
A576*            & yes (1,91)                  & 5.47 $\pm$ 0.68 $\times$ 10$^{-4}$ 	           & no         & yes(36,48,73)         & B55, HIFLUGCS 			 \\
PKS 0745-191    & yes(8)*                   & 5.53 $\pm$ 0.22 						           & yes(109)    & \ldots               & B55 						 \\ 
A644            & no(2)                        & 6.0 $\times$ 10$^{-4}$ 	           & no         & yes(13)                & B55 						 \\
A754            & yes(70)$^\dagger$          & 1.02 $\times$ 10$^{-3}$ 				           & no         & yes(48,62,64,70,83)         & B55, HIFLUGCS 			 \\
Hydra A         & yes(11,26,80,130)*        & 30.1 $\pm$ 0.3 						           & yes(109)    & \ldots               & B55, HIFLUGCS 					 \\
A1060*           & yes(82)*                  & 3.42 $\pm$ 0.34 $\times$ 10$^{-5}$ 	           & yes(143)    & yes(61)               & B55, HIFLUGCS 					 \\
A1367           & no(2)                        & 1.6 $\times$ 10$^{-4}$ 				       & no         & yes(25,48)            & B55, HIFLUGCS 				 \\
MKW 4           & yes(91)                   & 2.13 $\pm$ 0.45 $\times$ 10$^{-4}$ 	           & no         & \ldots               & HIFLUGCS 						 \\
ZwCl 1215       & no(1)                        & 4.79 $\times$ 10$^{-3}$ 				   & no         & \ldots               & HIFLUGCS 						 \\
M87             & yes(11,26,100)*            & 5.79 $\pm$ 0.24 $\times$ 10$^{-1}$ 	           & yes(109)    & yes(113)               & B55 						 \\
NGC 4636        & yes(7,34,51)*             & 2.82 $\pm$ 0.03 $\times$ 10$^{-4}$ 	           & yes(7)     & \ldots               & HIFLUGCS 						 \\
Centaurus       & yes(11,39)*               & 1.12 $\pm$ 0.03 $\times$ 10$^{-1}$ 	           & yes(109) & yes(48)               & B55, HIFLUGCS 					 \\
A1644           & yes(2)                    & 4.76 $\pm$ 0.15 $\times$ 10$^{-2}$ 	           & no         & yes(48,67,111)         & B55, HIFLUGCS 			 \\
A3532*           & yes(56,136)*$^\dagger$    & 7.66 $\pm$ 0.30 $\times$ 10$^{-1}$ 	           & yes(143)    & yes(136)               & B55 						 \\
A1650           & yes(55)                   & 7.7 $\pm$ 1.6 $\times$ 10$^{-4}$ 	               & no         & \ldots               & B55, HIFLUGCS 					 \\
A1651           & yes(2,91)                   & 1.32 $\pm$ 0.18 $\times$ 10$^{-2}$ 	           & no         & \ldots               & B55, HIFLUGCS 					 \\
Coma*            & yes(1)$^\dagger$          & 2.43 $\pm$ 0.01 $\times$ 10$^{-2}$ 	           & no         & yes(12,16,27,32,54)  & B55, HIFLUGCS 		 	\\
A1689           & yes(1)                    & 7.77 $\pm$ 1.53 $\times$ 10$^{-2}$ 	           & no         & yes(5,81,134)                & B55 						 \\
NGC 5044        & yes(28,51)*               & 5.95 $\pm$ 0.50 $\times$ 10$^{-4}$ 	           & yes(28)    & \ldots               & HIFLUGCS 						 \\
A1736           & yes(4)                    & 2.92 $\pm$ 0.35 $\times$ 10$^{-3}$	           & no         &\ldots                & B55, HIFLUGCS 					 \\
A3558*           & yes(2,91)$^\dagger$         & 2.11 $\pm$ 0.33 $\times$ 10$^{-3}$ 	           & no         & yes(48,135)            & B55, HIFLUGCS 				 \\
A3562           & yes(89)*$^\dagger$        & 2.8 $\pm$ 0.5 $\times$ 10$^{-4}$ 	               & no         & yes(48,49,135)         & B55, HIFLUGCS 			 \\
A3571           & yes(137)$^\dagger$        & 1.39 $\pm$ 0.06 $\times$ 10$^{-3}$ 	           & no         & yes(137)               & B55, HIFLUGCS 					 \\ 
A1795           & yes(11,15,46)*            & 8.45 $\pm$ 0.24 $\times$ 10$^{-1}$ 	           & yes(109)    & yes(48,86,133)            & B55, HIFLUGCS 				 \\
\hline
\end{tabular}
\end{minipage}
\end{table*}

\begin{table*}
\begin{minipage}{168mm}
\contcaption{}
\label{radio_info_table}
\begin{tabular}{@{}lccccccccccc} 
System$^a$& Radio$^b$ & $L_{\rm 1.4GHz}$$^c$ & Bubbles & Merger &  Sample$^d$ \\
\hline

PKS 1404-267    & yes(66)*                  & 7.45 $\pm$ 0.26 $\times$ 10$^{-2}$ 	           & yes(109)    & \ldots               & HIFLUGCS 						 \\
MKW 8*           & yes(1,91)                   & 4.02 $\pm$ 0.25 $\times$ 10$^{-4}$ 	           & no         & \ldots               & HIFLUGCS 						 \\
RXCJ 1504.1-0248& yes(2,91)$^\dagger$$^\dagger$         & 7.74 $\pm$ 0.27 $\times$ 10$^{-1}$ 	   & no         & yes(50)               & HIFLUGCS 						 \\
A2029           & yes(15,55,131)*$^\dagger$$^\dagger$ & 1.06 $\pm$ 0.24 					   & yes(109)    & yes(19)                & B55, HIFLUGCS 					 \\
A2052           & yes(11,15,138)*           & 1.56 $\pm$ 0.028 					               & yes(109)    & \ldots               & B55, HIFLUGCS 					 \\
MKW 3s          & yes(11)*                  & 6.780 $\pm$ 0.001 $\times$ 10$^{-2}$             & yes(109)    & yes(76)               & B55, HIFLUGCS 					 \\
A2065           & yes(1)                    & 1.20 $\pm$ 0.22 $\times$ 10$^{-2}$ 	           & no         & yes(23,48)            & B55, HIFLUGCS 				 \\
A2063           & yes(91,99)*               & 4.29 $\pm$ 0.28 $\times$ 10$^{-3}$ 	           & no         & yes(76)               & B55, HIFLUGCS 					 \\
A2142           & yes(91)$^\dagger$         & 4.60 $\pm$ 0.071 $\times$ 10$^{-3}$    & no         & yes(59,84,102,103,133)               & B55, HIFLUGCS 					 \\
A2147           & yes(91,99)*               & 4.37 $\pm$ 0.31 $\times$ 10$^{-3}$ 	           & no         & yes(77)               & B55, HIFLUGCS 					 \\
A2163           & no(2)$^\dagger$              & 1.74 $\times$ 10$^{-2}$				       & no         & yes(41,42,85,102)            & HIFLUGCS 					 \\
A2199           & yes(11,47,53)*$^\dagger$$^\dagger$  & 7.37 $\pm$ 0.16 $\times$ 10$^{-1}$ 	   & yes(109)    & yes(48,77)            & B55, HIFLUGCS 				 \\
A2204           & yes(119)*                  & 4.27 $\pm$ 0.15 $\times$ 10$^{-1}$ 	           & yes(119)    & yes(120)               & B55, HIFLUGCS 					 \\
Triangulum Aus.     & yes(3)                       & 4.6 $\pm$ 0.8 $\times$ 10$^{-3}$ 				       & no         & \ldots               & B55 						 \\
A2244           & yes(60,91)                & 6.77 $\pm$ 0.82 $\times$ 10$^{-3}$ 	           & no         & \ldots               & B55, HIFLUGCS 					 \\
A2256           & no(2)$^\dagger$          & 1.18 $\times$ 10$^{-3}$ 			               & no         & yes(10,21,48,69,87,127)   & B55, HIFLUGCS		  \\
Ophiuchus       & yes(2,4,93)$^\dagger$$^\dagger$     & 5.17 $\pm$ 0.18 $\times$ 10$^{-3}$ 	   & no         & yes(90)               & B55 						 \\
A2255           & no(1)$^\dagger$              & 2.34 $\times$ 10$^{-3}$ 				   & no         & yes(88,54,108,117)             & B55, HIFLUGCS 				 \\
A2319           & no(2)$^\dagger$          & 3.7 $\times$ 10$^{-4}$ 	           & no         & yes(48,54,96)            & B55 					 \\
Cygnus A        & yes(11,17)*               & 1092 $\pm$ 45 						           & yes(109)    & \ldots               & B55 						 \\
A3667           & no(2)$^\dagger$              & 1.29 $\times$ 10$^{-3}$ 				       & no         & yes(18,44,48,101,102,115,140)            & B55, HIFLUGCS 				 \\
Sersic 159/03   & yes(11)*                  & 1.92 $\pm$ 0.20 $\times$ 10$^{-1}$ 	           & yes(109)    & probable(29)          & HIFLUGCS 					 \\
A2589*           & no(2)                        & 5.83 $\times$ 10$^{-4}$ 				       & no         & \ldots               & HIFLUGCS 						 \\
A2597           & yes(11,19,123)*           & 3.32 $\pm$ 0.09 						           & yes(109)    & \ldots               & B55, HIFLUGCS 					 \\
A2634*           & yes(15,57,58)*            & 1.730 $\pm$ 0.086 					           & no         & \ldots               & HIFLUGCS 						 \\
A2657           & no(2)                        & 5.49 $\times$ 10$^{-4}$ 				       & no         & \ldots               & HIFLUGCS					 \\
A4038*           & yes(125)$^\dagger$        & 5.19 $\pm$ 0.08 $\times$ 10$^{-3}$ 	           & no         & probable(14,125)          & B55,HIFLUGCS 				 \\
\hline
\end{tabular}
\end{minipage}
\end{table*}

\begin{table*}
\begin{minipage}{168mm}
\contcaption{}
\label{radio_info_table}
\begin{tabular}{@{}lccccccccccc} 
System$^a$& Radio$^b$ & $L_{\rm 1.4GHz}$$^c$ & Bubbles & Merger &  Sample$^d$ \\
\hline

A4059           & yes(11,131)*              & 6.82 $\pm$ 0.23 $\times$ 10$^{-1}$ 	           & yes(109)    & yes(112)               & B55, HIFLUGCS 					 \\       
\hline
\end{tabular}
 \label{radio_info_table}
 \medskip
References: (1) FIRST - Faint Images of the Radio Sky at Twenty-cm \citep{beck95}; (2) NVSS - The NRAO VLA Sky Survey at 1.4 GHz \citep{cond98}; (3) SUMSS - The Sydney University Molonglo Sky Survey at 843 MHz \citep{bock99};  (4) TGSS (TIFR GMRT Sky Survey) by S. K. Sirothia, N. G. Kantharia, C. H. Ishwara-Chandra \& Gopal-Krishna; (5) \citet{ande04}; (6) \citet{bagc06}; (7) \citet{bald09}; (8) \citet{baum91}; (9) \citet{bels05}; (10) \citet{berr02}; (11) \citet{birz08}; (12) \citet{brav00}; (13) \citet{buot05}; (14) \citet{burg04}; (15) \citet{burn90}; (16) \citet{burn94}; (17) \citet{cari91}; (18) \citet{carr12}; (19) \citet{clar04a}; (20) \citet{clar05}; (21) \citet{clar06a}; (22) \citet{clar09}; (23) \citet{chat06}; (24) \citet{chur03}; (25) \citet{cort04}; (26) \citet{cott09}; (27) \citet{davi93}; (28) \citet{davi10}; (29) \citet{dePl06}; (30) \citet{deRu86}; (31) \citet{dong10}; (32) \citet{donn99}; (33) \citet{donn01}; (34) \citet{dunn10}; (35) \citet{dupk07}; (36) \citet{dupk07a}; (37) \citet{durr05}; (38) \citet{fabi03}; (39) \citet{fabi05}; (40) \citet{fere91}; (41) \citet{fere01}; (42) \citet{fere04}; (43) \citet{fino06}; (44) \citet{fino10}; (45) \citet{gast09}; (46) \citet{ge93}; (47) \citet{gent07}; (48) \citet{ghiz10}; (49) \citet{giac05}; (50) \citet{giac11}; (51) \citet{giac11b}; (52) \citet{giov93}; (53) \citet{giov98a}; (54) \citet{govo01}; (55) \citet{govo09}; (56) \citet{greg94}; (57) \citet{hard04}; (58) \citet{hard05}; (59) \citet{harr77}; (60) \citet{hani84}; (61) \citet{haya04}; (62) \citet{henr93}; (63) \citet{henr02}; (64) \citet{henr04}; (65) \citet{huds06}; (66) \citet{john05}; (67) \citet{john10}; (68) \citet{joHo08}; (69) \citet{kale10}; (70) \citet{kass01}; (71) \citet{kawa09}; (72) \citet{kemp02}; (73) \citet{kemp04}; (74) \citet{kim95}; (75) \citet{kraw03}; (76) \citet{krem99}; (77) \citet{krem02}; (78) \citet{VirL04}; (79) \citet{lane02}; (80) \citet{lane04}; (81) \citet{leon11}; (82) \citet{lind85};  (83) \citet{maca10}; (84) \citet{mark00}; (85) \citet{mark01}; (86)  \citet{mark01b}; (87) \citet{mill03}; (88) \citet{mill03a}; (89) \citet{mill05}; (90) \citet{mill10}; (91) \citet{mitt09}; (92) \citet{morg99}; (93) \citet{murg10}; (94) \citet{murg10a}; (95) \citet{murg11}; (96) \citet{OHar04}; (97) \citet{otan98}; (98) \citet{owen85}; (99) \citet{owen97}; (100) \citet{owen00}; (101) \citet{ower09}; (102) \citet{ower09b}; (103) \citet{ower11}; (104) \citet{paol03}; (105) \citet{parm86}; (106) \citet{paul11}; (107) \citet{pele90}; (108) \citet{pizz09}; (109) \citet{raff06}; (110) \citet{rand10}; (111) \citet{reip04}; (112) \citet{reyn08}; (113) \citet{roed11}; (114) \citet{roed12}; (115) \citet{rott97}; (116) \citet{sake04}; (117) \citet{sake06}; (118) \citet{sand09a}; (119) \citet{sand09}; (120)  \citet{sand05}; (121) \citet{sauv05}; (122) \citet{sara95b};  (123) \citet{sara95a}; (124) \citet{shur08}; (125) \citet{slee98}; (126) \citet{slee01}; (127) \citet{sun02}; (128) \citet{taki03}; (129) \citet{tana10}; (130) \citet{tayl90}; (131) \citet{tayl94}; (132) \citet{titt01}; (133) \citet{titt05}; (134) \citet{vacc11}; (135) \citet{vent00}; (136) \citet{vent01}; (137) \citet{vent02}; (138) \citet{vent04}; (139) \citet{vikh97}; (140) \citet{vikh01b}; (141) \citet{wang10a}; (142) \citet{wern06}; (143) This work.\\
$^a$The asterisk indicates the corona class systems (see Section \ref{comp_samples}).\\
$^b$The presence of a central radio source is based on information from the literature (the references are given in parentheses). The asterisk indicates the presence of radio lobes associated with the central radio source. The $^\dagger$ symbol indicates the presence of diffuse radio emission, such as halos or merger relics: A399, A401 \citep{murg10a}, A3376 \citep{bagc06}, A754 \citep{kass01,bacc03,kale09,maca10}, A3532 \citep{vent01}, Coma \citep{kim89,deis97,govo01}, A1689 \citep{vacc11}, A3558 \citep{vent00}, A3562 \citep{vent00,vent03,giac05}, A3571 \citep{vent02}, A2142 \citep{giov00,harr77}, A2163 \citep{fere01}, A2256 \citep{clar06,clar06a,vanW09,kale10}, A2255 \citep{govo05,pizz09,bren08}, A2319 \citep{govo01}, A3667 \citep{rott97,john03,carr12}, A4038 \citep{slee98}. The  $^\dagger$$^\dagger$ symbol indicates the presence of mini-haloes: A85 \citep{slee01}, A133 \citep{slee01,birz08,rand10}, RXCJ 1504.1-0248 \citep{giac11}, A2029 \citep{govo09}, A2199 \citep{bacc03}, Ophiuchus \citep{govo09,murg10}.\\
$^c$In units of 10$^{32}$ erg s$^{-1}$ Hz$^{-1}$. The radio luminosities are from \citet{mitt09} where available. The numbers without errors are upper limits derived from the noise in the NVSS images. For Ophiuchus the NVSS flux was used to calculate the radio luminosity; for Triangulum Aus. the 1.4 GHz flux was extrapolated from the SUMSS survey flux ($S_{875}=12 \pm 2$ mJy);  for 3C129.1 the flux and the spectral index are from \citet{VirL04};  for Persues, M87 and Cygnus A the fluxes and spectral indices are from \citet{birz08}; for A1060 the 1.4 GHz flux was extrapolated from the 4.5 GHz one \citep{lind85}; for A1650  the flux is from \citet{govo09}; for A1689 the flux is from the FIRST survey; for A1736 the TGSS survey at 150 MHz was used ($S_{150}=55.3 \pm 6.6$ mJy); for PKS 0745-191 the flux and spectral index are from \citet{baum91}; for A3532 the flux and spectral index are from \citet{vent01}. For A644, A2319 and AWM 7 the radio luminosities are upper limits from the NVSS images (however, for A2319 the NVSS image shows a radio source close to the BCG location). When not available a spectral index of $-1$ was adopted.\\
$^d$Complete flux limited X-ray samples: B55 (the Brightest 55 clusters of galaxies), and HIFLUGCS (the HIghest X-ray FLUx Galaxy Cluster Sample). See Section \ref{comp_samples} for details.
\end{minipage}
\end{table*}

\end{document}